\documentclass[12pt,a4paper]{article}
\usepackage[latin1]{inputenc}
\usepackage[T1]{fontenc}
\usepackage{amsmath}
\usepackage{amsfonts}
\usepackage{amssymb}
\usepackage[english]{babel}
\usepackage[dvips]{graphicx,color}

\def\graph#1{
\begin{array}{c}\mbox{
\includegraphics*[height=1cm]{#1.eps}
}\end{array}
             }
\begin{document}
\newcommand {\nn}{\nonumber \\}
\newcommand {\Tr}{{\rm Tr\,}}
\newcommand {\tr}{{\rm tr\,}}
\newcommand {\e}{{\rm e}}
\newcommand {\etal}{{\it et al.}}
\newcommand {\m}{\mu}
\newcommand {\n}{\nu}
\newcommand {\pl}{\partial}
\newcommand {\p} {\phi}
\newcommand {\vp}{\varphi}
\newcommand {\vpc}{\varphi_c}
\newcommand {\al}{\alpha}
\newcommand {\be}{\beta}
\newcommand {\ga}{\gamma}
\newcommand {\Ga}{\Gamma}
\newcommand {\x}{\xi}
\newcommand {\ka}{\kappa}
\newcommand {\la}{\lambda}
\newcommand {\La}{\Lambda}
\newcommand {\si}{\sigma}
\newcommand {\Si}{\Sigma}
\newcommand {\sh}{\theta}
\newcommand {\Th}{\Theta}
\newcommand {\om}{\omega}
\newcommand {\Om}{\Omega}
\newcommand {\ep}{\epsilon}
\newcommand {\vep}{\varepsilon}
\newcommand {\na}{\nabla}
\newcommand {\del}  {\delta}
\newcommand {\Del}  {\Delta}
\newcommand {\mn}{{\mu\nu}}
\newcommand {\ls}   {{\lambda\sigma}}
\newcommand {\ab}   {{\alpha\beta}}
\newcommand {\gd}   {{\gamma\delta}}
\newcommand {\half}{ {\frac{1}{2}} }
\newcommand {\third}{ {\frac{1}{3}} }
\newcommand {\fourth} {\frac{1}{4} }
\newcommand {\sixth} {\frac{1}{6} }
\newcommand {\sqtwo} {\sqrt{2}}
\newcommand {\sqg} {\sqrt{g}}
\newcommand {\fg}  {\sqrt[4]{g}}
\newcommand {\invfg}  {\frac{1}{\sqrt[4]{g}}}
\newcommand {\sqZ} {\sqrt{Z}}
\newcommand {\sqk} {\sqrt{\kappa}}
\newcommand {\sqt} {\sqrt{t}}
\newcommand {\sql} {\sqrt{l}}
\newcommand {\reg} {\frac{1}{\epsilon}}
\newcommand {\fpisq} {(4\pi)^2}
\newcommand {\Acal}{{\cal A}}
\newcommand {\Lcal}{{\cal L}}
\newcommand {\Ocal}{{\cal O}}
\newcommand {\Dcal}{{\cal D}}
\newcommand {\Ncal}{{\cal N}}
\newcommand {\Mcal}{{\cal M}}
\newcommand {\scal}{{\cal s}}
\newcommand {\Dvec}{{\hat D}}   
\newcommand {\dvec}{{\vec d}}
\newcommand {\Evec}{{\vec E}}
\newcommand {\Hvec}{{\vec H}}
\newcommand {\Vvec}{{\vec V}}
\newcommand {\lsim}{
\vbox{\baselineskip=4pt \lineskiplimit=0pt \kern0pt 
\hbox{$<$}\hbox{$\sim$}}
                    }
\newcommand {\rpl}{{\vec \partial}}
\def\overleftarrow#1{\vbox{\ialign{##\crcr
 $\leftarrow$\crcr\noalign{\kern-1pt\nointerlineskip}
 $\hfil\displaystyle{#1}\hfil$\crcr}}}
\def\lpl{{\overleftarrow\partial}}
\newcommand {\Btil}{{\tilde B}}
\newcommand {\atil}{{\tilde a}}
\newcommand {\btil}{{\tilde b}}
\newcommand {\ctil}{{\tilde c}}
\newcommand {\dtil}{{\tilde d}}
\newcommand {\Ftil}{{\tilde F}}
\newcommand {\Ktil}  {{\tilde K}}
\newcommand {\Ltil}  {{\tilde L}}
\newcommand {\mtil}{{\tilde m}}
\newcommand {\ttil} {{\tilde t}}
\newcommand {\Qtil}  {{\tilde Q}}
\newcommand {\Rtil}  {{\tilde R}}
\newcommand {\Stil}{{\tilde S}}
\newcommand {\Ztil}{{\tilde Z}}
\newcommand {\altil}{{\tilde \alpha}}
\newcommand {\betil}{{\tilde \beta}}
\newcommand {\etatil} {{\tilde \eta}}
\newcommand {\latil}{{\tilde \lambda}}
\newcommand {\Latil}{{\tilde \Lambda}}
\newcommand {\ptil}{{\tilde \phi}}
\newcommand {\Ptil}{{\tilde \Phi}}
\newcommand {\natil} {{\tilde \nabla}}
\newcommand {\xitil} {{\tilde \xi}}
\newcommand {\Hbtil} {{\Huge {\~{b}}}}
\newcommand {\Hctil} {{\Huge {\~{c}}}}
\newcommand {\Hdtil} {{\Huge {\~{d}}}}
\newcommand {\Ahat}{{\hat A}}
\newcommand {\ahat}{{\hat a}}
\newcommand {\Rhat}{{\hat R}}
\newcommand {\Shat}{{\hat S}}
\newcommand {\ehat}{{\hat e}}
\newcommand {\mhat}{{\hat m}}
\newcommand {\shat}{{\hat s}}
\newcommand {\Dhat}{{\hat D}}   
\newcommand {\Vhat}{{\hat V}}   
\newcommand {\xhat}{{\hat x}}
\newcommand {\Zhat}{{\hat Z}}
\newcommand {\Gahat}{{\hat \Gamma}}
\newcommand {\Phihat} {{\hat \Phi}}
\newcommand {\phihat} {{\hat \phi}}
\newcommand {\vphat} {{\hat \varphi}}
\newcommand {\nah} {{\hat \nabla}}
\newcommand {\etahat} {{\hat \eta}}
\newcommand {\omhat} {{\hat \omega}}
\newcommand {\psihat} {{\hat \psi}}
\newcommand {\thhat} {{\hat \theta}}
\newcommand {\gh}  {{\hat g}}
\newcommand {\abar}{{\bar a}}
\newcommand {\Abar}{{\bar A}}
\newcommand {\cbar}{{\bar c}}
\newcommand {\bbar}{{\bar b}}
\newcommand {\gbar}{\bar{g}}
\newcommand {\Bbar}{{\bar B}}
\newcommand {\Dbar}{{\bar D}}
\newcommand {\fbar}{{\bar f}}
\newcommand {\Fbar}{{\bar F}}
\newcommand {\kbar}  {{\bar k}}
\newcommand {\Kbar}  {{\bar K}}
\newcommand {\Lbar}  {{\bar L}}
\newcommand {\Qbar}  {{\bar Q}}
\newcommand {\Wbar}  {{\bar W}}
\newcommand {\albar}{{\bar \alpha}}
\newcommand {\bebar}{{\bar \beta}}
\newcommand {\epbar}{{\bar \epsilon}}
\newcommand {\labar}{{\bar \lambda}}
\newcommand {\psibar}{{\bar \psi}}
\newcommand {\vpbar}{{\bar \varphi}}
\newcommand {\Psibar}{{\bar \Psi}}
\newcommand {\Phibar}{{\bar \Phi}}
\newcommand {\chibar}{{\bar \chi}}
\newcommand {\sibar}{{\bar \sigma}}
\newcommand {\xibar}{{\bar \xi}}
\newcommand {\thbar}{{\bar \theta}}
\newcommand {\Thbar}{{\bar \Theta}}
\newcommand {\bbartil}{{\tilde {\bar b}}}
\newcommand {\aldot}{{\dot{\alpha}}}
\newcommand {\bedot}{{\dot{\beta}}}
\newcommand {\gadot}{{\dot{\gamma}}}
\newcommand {\deldot}{{\dot{\delta}}}
\newcommand {\alp}{{\alpha'}}
\newcommand {\bep}{{\beta'}}
\newcommand {\gap}{{\gamma'}}
\newcommand {\bfZ} {{\bf Z}}
\newcommand {\BFd} {{\bf d}}
\newcommand  {\vz}{{v_0}}
\newcommand  {\ez}{{e_0}}
\newcommand  {\mz}{{m_0}}
\newcommand  {\xf}{{x^5}}
\newcommand  {\yf}{{y^5}}
\newcommand  {\Zt}{{Z$_2$}}
\newcommand {\intfx} {{\int d^4x}}
\newcommand {\intdX} {{\int d^5X}}
\newcommand {\inttx} {{\int d^2x}}
\newcommand {\change} {\leftrightarrow}
\newcommand {\ra} {\rightarrow}
\newcommand {\larrow} {\leftarrow}
\newcommand {\ul}   {\underline}
\newcommand {\pr}   {{\quad .}}
\newcommand {\com}  {{\quad ,}}
\newcommand {\q}    {\quad}
\newcommand {\qq}   {\quad\quad}
\newcommand {\qqq}   {\quad\quad\quad}
\newcommand {\qqqq}   {\quad\quad\quad\quad}
\newcommand {\qqqqq}   {\quad\quad\quad\quad\quad}
\newcommand {\qqqqqq}   {\quad\quad\quad\quad\quad\quad}
\newcommand {\qqqqqqq}   {\quad\quad\quad\quad\quad\quad\quad}
\newcommand {\lb}    {\linebreak}
\newcommand {\nl}    {\newline}

\newcommand {\vs}[1]  { \vspace*{#1 cm} }

\newcommand {\MPL}  {Mod.Phys.Lett.}
\newcommand {\NP}   {Nucl.Phys.}
\newcommand {\PL}   {Phys.Lett.}
\newcommand {\PR}   {Phys.Rev.}
\newcommand {\PRL}   {Phys.Rev.Lett.}
\newcommand {\IJMP}  {Int.Jour.Mod.Phys.}
\newcommand {\CMP}  {Commun.Math.Phys.}
\newcommand {\JMP}  {Jour.Math.Phys.}
\newcommand {\AP}   {Ann.of Phys.}
\newcommand {\PTP}  {Prog.Theor.Phys.}
\newcommand {\NC}   {Nuovo Cim.}
\newcommand {\CQG}  {Class.Quantum.Grav.}


\newcommand {\npl}  {{\frac{n\pi}{l}}}
\newcommand {\mpl}  {{\frac{m\pi}{l}}}
\newcommand {\kpl}  {{\frac{k\pi}{l}}}

\def\ocirc#1{#1^{^{{\hbox{\smallr\llap{o}}}}}}
\def\ogamma{\ocirc{\gamma}{}}
\def\oM{{\buildrel {\hbox{\smallr{o}}} \over M}}
\def\osigma{\ocirc{\sigma}{}}

\def\overleftrightarrow#1{\vbox{\ialign{##\crcr
 $\leftrightarrow$\crcr\noalign{\kern-1pt\nointerlineskip}
 $\hfil\displaystyle{#1}\hfil$\crcr}}}
\def\overnab{{\overleftrightarrow\nabslash}}

\def\va{{a}}
\def\vb{{b}}
\def\vc{{c}}
\def\tilpsi{{\tilde\psi}}
\def\tbpsi{{\tilde{\bar\psi}}}

\newcommand {\sqxx}  {\sqrt {x^2+1}}   
\newcommand {\gago}  {\gamma^5}
\newcommand {\Pp}  {P_+}
\newcommand {\Pm}  {P_-}
\newcommand {\GfMp}  {G^{5M}_+}
\newcommand {\GfMpm}  {G^{5M'}_-}
\newcommand {\GfMm}  {G^{5M}_-}
\newcommand {\Omp}  {\Omega_+}    
\newcommand {\Omm}  {\Omega_-}
\def\Aslash{{}\hbox{\hskip2pt\vtop
 {\baselineskip23pt\hbox{}\vskip-24pt\hbox{/}}
 \hskip-11.5pt $A$}}
\def\Rslash{{}\hbox{\hskip2pt\vtop
 {\baselineskip23pt\hbox{}\vskip-24pt\hbox{/}}
 \hskip-11.5pt $R$}}

\def\kslash{
{}\hbox       {\hskip2pt\vtop
                   {\baselineskip23pt\hbox{}\vskip-24pt\hbox{/}}
               \hskip-8.5pt $k$}
           }    
\def\qslash{
{}\hbox       {\hskip2pt\vtop
                   {\baselineskip23pt\hbox{}\vskip-24pt\hbox{/}}
               \hskip-8.5pt $q$}
           }    
\def\dslash{
{}\hbox       {\hskip2pt\vtop
                   {\baselineskip23pt\hbox{}\vskip-24pt\hbox{/}}
               \hskip-8.5pt $\partial$}
           }    
\def\dbslash{{}\hbox{\hskip2pt\vtop
 {\baselineskip23pt\hbox{}\vskip-24pt\hbox{$\backslash$}}
 \hskip-11.5pt $\partial$}}

\def\Kbslash{{}\hbox{\hskip2pt\vtop
 {\baselineskip23pt\hbox{}\vskip-24pt\hbox{$\backslash$}}
 \hskip-11.5pt $K$}}
\def\Ktilbslash{{}\hbox{\hskip2pt\vtop
 {\baselineskip23pt\hbox{}\vskip-24pt\hbox{$\backslash$}}
 \hskip-11.5pt ${\tilde K}$}}
\def\Ltilbslash{{}\hbox{\hskip2pt\vtop
 {\baselineskip23pt\hbox{}\vskip-24pt\hbox{$\backslash$}}
 \hskip-11.5pt ${\tilde L}$}}
\def\Qtilbslash{{}\hbox{\hskip2pt\vtop
 {\baselineskip23pt\hbox{}\vskip-24pt\hbox{$\backslash$}}
 \hskip-11.5pt ${\tilde Q}$}}
\def\Rtilbslash{{}\hbox{\hskip2pt\vtop
 {\baselineskip23pt\hbox{}\vskip-24pt\hbox{$\backslash$}}
 \hskip-11.5pt ${\tilde R}$}}
\def\Kbarbslash{{}\hbox{\hskip2pt\vtop
 {\baselineskip23pt\hbox{}\vskip-24pt\hbox{$\backslash$}}
 \hskip-11.5pt ${\bar K}$}}
\def\Lbarbslash{{}\hbox{\hskip2pt\vtop
 {\baselineskip23pt\hbox{}\vskip-24pt\hbox{$\backslash$}}
 \hskip-11.5pt ${\bar L}$}}
\def\Rbarbslash{{}\hbox{\hskip2pt\vtop
 {\baselineskip23pt\hbox{}\vskip-24pt\hbox{$\backslash$}}
 \hskip-11.5pt ${\bar R}$}}
\def\Qbarbslash{{}\hbox{\hskip2pt\vtop
 {\baselineskip23pt\hbox{}\vskip-24pt\hbox{$\backslash$}}
 \hskip-11.5pt ${\bar Q}$}}
\def\Acalbslash{{}\hbox{\hskip2pt\vtop
 {\baselineskip23pt\hbox{}\vskip-24pt\hbox{$\backslash$}}
 \hskip-11.5pt ${\cal A}$}}


\begin{flushright}
Martch 2006\\
UWThPh-2006-8\\
hep-th/0603220 \\
\end{flushright}
\title{Graphical Representation of SUSY and C-Program Calculation
      }
\author{Shoichi ICHINOSE
\footnote{
On leave of absence from 
Laboratory of Physics, 
School of Food and Nutritional Sciences,
University of Shizuoka.
Untill 31 March, 2006.
\newline 
E-mail address:\ ichinose@u-shizuoka-ken.ac.jp
                  }
\\
  {\normalsize\it Institut f{\"u}r Theoretische Physik, 
Universit{\"a}t Wien}\\
  {\normalsize\it Boltzmanngasse 5, A-1090 Vienna, Austria}
       }
%
\date{}
\maketitle
\begin{abstract}
\noindent 
We present a graphical representation of the supersymmetry
and a C-program for the graphical calculation. 
Calculation is demonstrated for 4D Wess-Zumino model and for
Super QED.
The chiral operators are graphically expressed
in an illuminating way. The tedious part of SUSY calculation, due to
manipulating chiral suffixes, reduces considerably. The application
is diverse.
\end{abstract}

PACS NO:\  
02.10.Ox, 
02.70.-c, 
02.90.+p, 
11.30.Pb, 
11.30.Rd 
\nl

\section{Introduction}\label{sec:intro}
The supersymmetry is the symmetry between fermions and bosons. 
It was introduced in the mid 70's. At present the experiment
does not yet confirm the symmetry, but everybody accepts its importance
in nature and expects fruitful results in the future developement. 
The requirement of such a high symmetry costs a sophisticated
structure which makes its dynamical analysis difficult. 
In this circumstance, we propose a calculational technique which
utilizes the graphical representation of SUSY. 
The representation was proposed in \cite{SI03,SUSY2004}.
\footnote{
An improved version of Ref.\cite{SI03} has recently appeared as Ref.\cite{SI06UW07}.
} 
The spinor is represented as a slanted line with a direction. Its chirality
is represented by the way the line is drawn. The introductory
explantion is given in the text. The advantage of the graph expression
is the use of the {\it graph indices}. Every independent graph, which corresponds
to a unique term in the ordinary calculation, is classified by a set of graph indices.
Hence the main efforts of programinng is devoted to find good graph indices
and to count them. SUSY calculation
generally is not a simple algebraic or combinatoric or analytical one. 
It involves the vast branch of mathematics including Grassmann algebra.
The delicate property of chirality is produced in this environment. 
Hence it seems that the ordinary(popular) programs, such as Mathematica, Maple, REDUCE, 
MAXIMA, form, do not work. It requires a more fundamental language.
We take C-language and present a first-step program. Future development
to the level of the previously cited ordinary programs is much expected.

The notation in the text is based on the standard textbook by Wess and Bagger\cite{WB92}.
\section{Spinors in SUSY:\ Graphical Representation and Storage Form}\label{sec:spinor}
Weyl spinors have the SU(2)$_L \times$SU(2)$_R$ structure.
The {\it chiral} suffix $\al$, appearing in $\psi^\al$ or $\psi_\al$, 
represents (fundamental representation, doublet representation)
SU(2)$_L$ and the {\it anti-chiral} suffix $\aldot$, appearing in
$\psibar^\aldot$ or $\psibar_\aldot$, represents SU(2)$_R$. 
The raising and lowering of suffixes are done
by the antisymmetric tensors $\ep^{\al\be}$ and $\ep_{\al\be}$.
\begin{eqnarray}
(\ep^{\al\be})=
\left(
\begin{array}{cc}
0 & 1 \\
-1 & 0
\end{array}
\right)\com\q
(\ep_{\aldot\bedot})=
\left(
\begin{array}{cc}
0 & -1 \\
1 & 0
\end{array}
\right)\com\q
\ep^{\al\be}\ep_{\be\ga}=\del^\al_\ga  \com\nn
\psi^\al=\ep^{\al\be}\psi_\be\com\q 
\psibar_\aldot=\ep_{\aldot\bedot}\psibar^\bedot
\pr
\label{int1}
\end{eqnarray}
They are graphically expressed by Fig.1. 
\begin{figure}[htbp]
\centerline{
\includegraphics*[height=5cm]{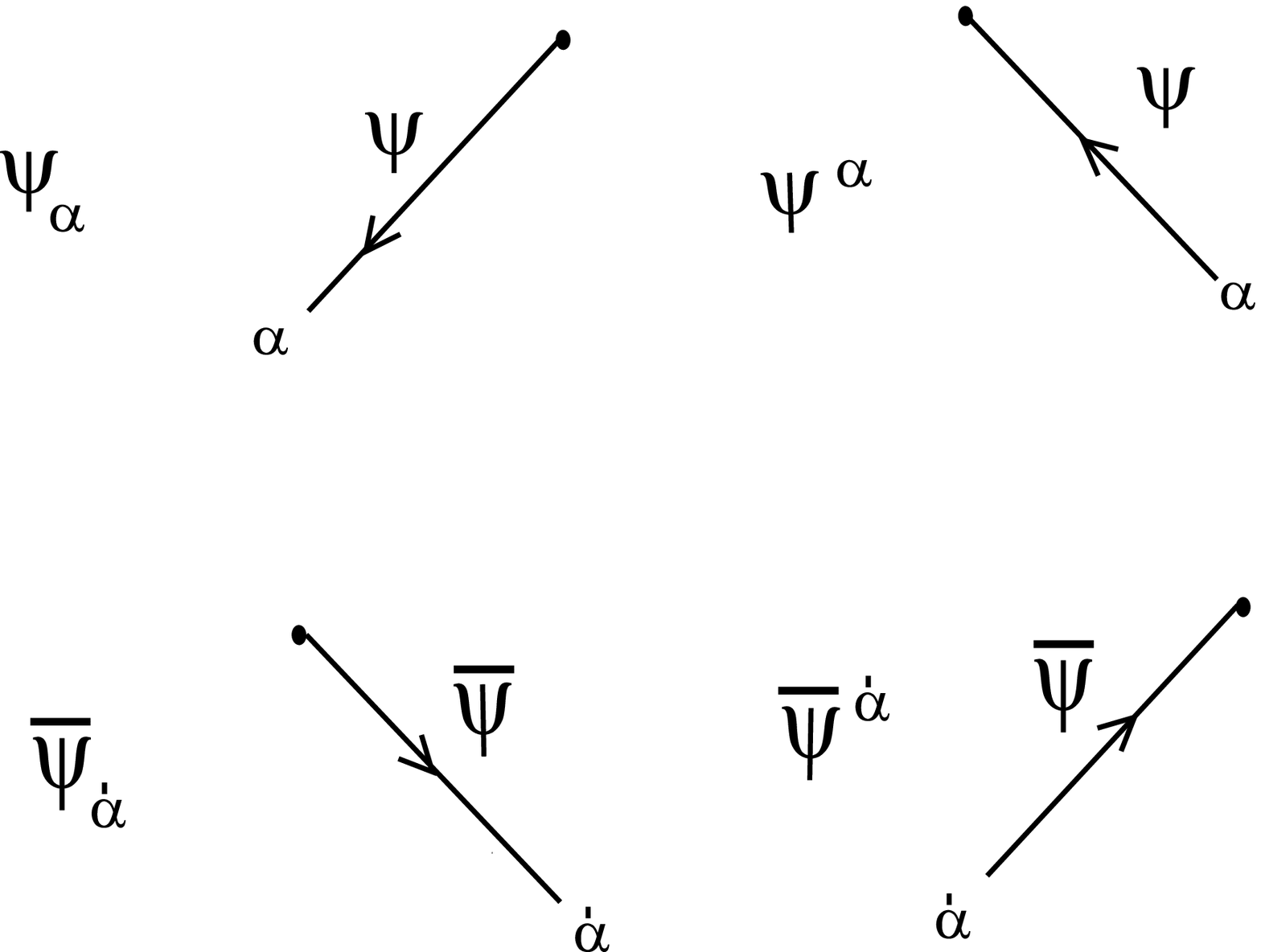}
           }
\label{LF1}
   \caption{
Weyl fermions.
   }
\end{figure}
We encode them as follows. We use 2 dimensional array
with the size 2$\times$2. The four chiral spinors 
are stored in C-program as the array psi[\ ][\ ].

\vspace{5mm}
{\bf Fermionic Fields}\nl
\vspace{3mm}
(1) (Weyl) Spinor [Symbol: p\ ;\ Dimension: M$^{3/2}$ ]\nl
\shortstack[l]{
$\psi^\al$\\
psi[0,0]=$\al$\\
psi[0,1]=empty\\
psi[1,0]=empty\\
psi[1,1]=empty
             }  
\q
\shortstack[l]{
$\psi_\al$\\
psi[0,0]=empty\\
psi[0,1]=$\al$\\
psi[1,0]=empty\\
psi[1,1]=empty
             }
\q
\shortstack[l]{
$\psibar_\aldot$\\
psi[0,0]=empty\\
psi[0,1]=empty\\
psi[1,0]=empty\\
psi[1,1]=$\aldot$
             }
\q
\shortstack[l]{
$\psibar^\aldot$\\
psi[0,0]=empty\\
psi[0,1]=empty\\
psi[1,0]=$\aldot$\\
psi[1,1]=empty
             }

The first column takes two numbers 0 and 1; 0 expresses a 'chiral' operator
$\psi$, while 1 expresses an 'anti-chiral' operator $\psibar$. The second
column also takes the two numbers; 0 expresses an 'up' suffix, while 1 expresses
an 'down' one. 

Note:\ Chiral spinor suffixes $\al,\be,\cdots$ are expressed
, in the present C-program, by positive {\it odd} number integers 1,3,$\cdots$, while
anti-chiral ones $\aldot,\bedot,\cdots$ are by positive {\it even} number integers
2,4,$\cdots$. This convension (discriminative use of even and odd integers) is, 
at this stage, 
rather redundant in the sense that the chirality ($\psi$ or $\psibar$)
can be read by the first element number of the array psi[2][2]
for a non-empty data, i.e. 0 for $\psi$ (psi[0][*]=$\al$)
and 1 for $\psibar$ (psi[1][*]=$\aldot$). 
(The situation is the same for some other spinors $\pl_m\psi, \sh, \cdots$. 
See the later discription. )
The convension, however, 
will soon become important to discriminate the chirality
of the spinor matrices; $\si$ and $\sibar$.

Note:\ 'empty' is expressed by a default number (, for example, 99) in the program.  \nl
\nl
(2) First derivative of spinor [Symbol: q\ ;\ Dimension: M$^{5/2}$ ]\nl
\q The first derivative of the spinor is graphically expressed by 
the upper graph of Fig.2.
\begin{figure}[htbp]
\centerline{
\includegraphics*[height=6cm]{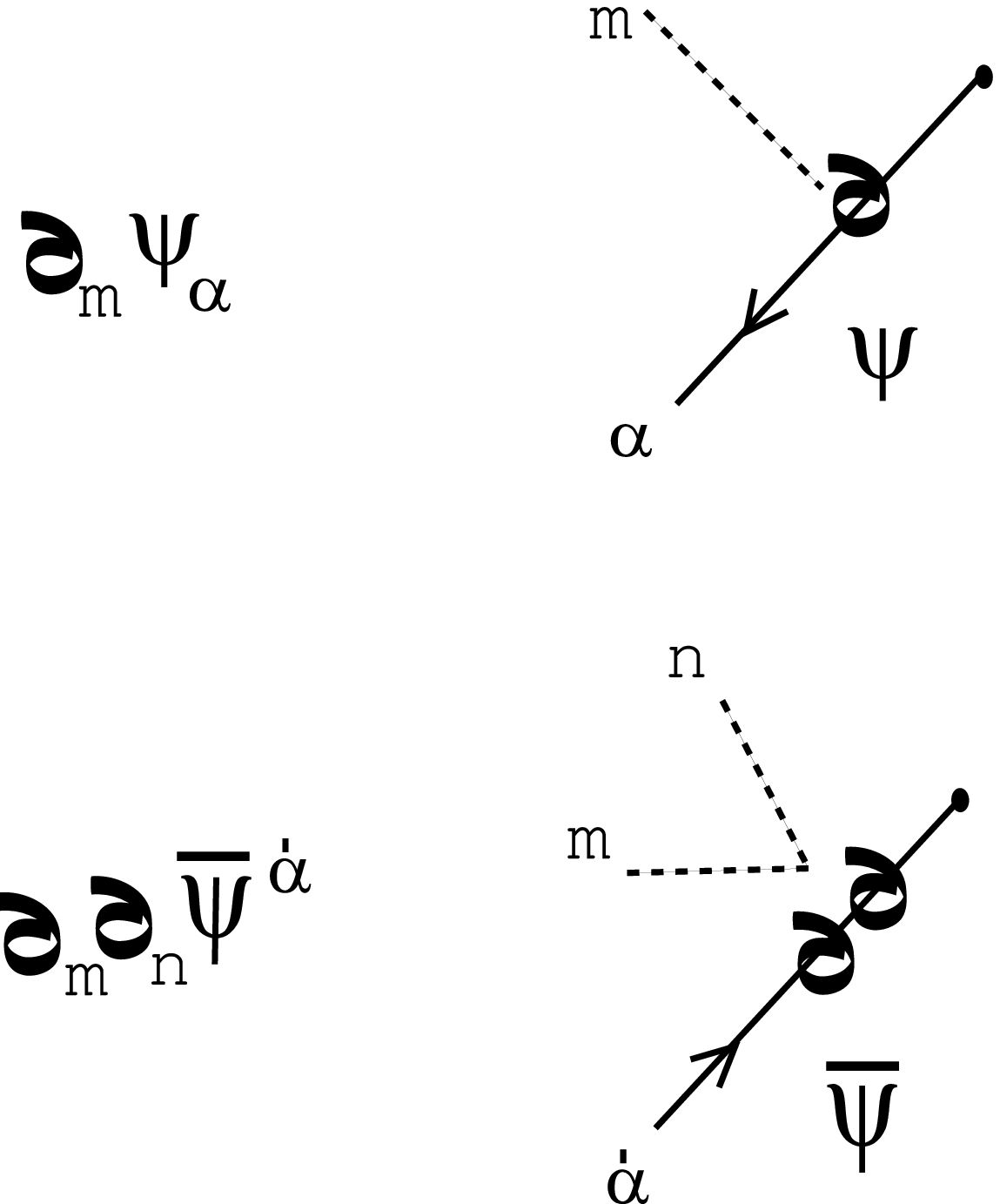}
           }
\label{LF2}
   \caption{
Derivatives of fermions.
   }
\end{figure}
It is stored by one 2$\times$2 array dps[\ ][\ ] and one variable dpsv.\nl
\shortstack[l]{
$\pl_m\psi^\al$\\
dps[0,0]=$\al$\\
dps[0,1]=empty\\
dps[1,0]=empty\\
dps[1,1]=empty\\
dpsv=m
             }
\q
\shortstack[l]{
$\pl_m\psi_\al$\\
dps[0,0]=empty\\
dps[0,1]=$\al$\\
dps[1,0]=empty\\
dps[1,1]=empty\\
dpsv=m
             }
\q
\shortstack[l]{
$\pl_m\psibar_\aldot$\\
dps[0,0]=empty\\
dps[0,1]=empty\\
dps[1,0]=empty\\
dps[1,1]=$\aldot$\\
dpsv=m
             }
\q
\shortstack[l]{
$\pl_m\psibar^\aldot$\\
dps[0,0]=empty\\
dps[0,1]=empty\\
dps[1,0]=$\aldot$\\
dps[1,1]=empty\\
dpsv=m
             }
\q

Here the vector suffix expresses the Lorents suffix of
a differetial operator.

Note: The vector suffixes m, n, $\cdots$ are expressed 
by 51, 52, $\cdots$ in the present program. \nl
\nl
(3) Sigma Matrix [Symbol: s\ ;\ Dimension: M$^0$ ]\nl
\q Sigma matrices $\si^m, \sibar^m$ are graphically expressed in Fig.3.\nl 
\begin{figure}[htbp]
\centerline{
\includegraphics*[height=8cm]{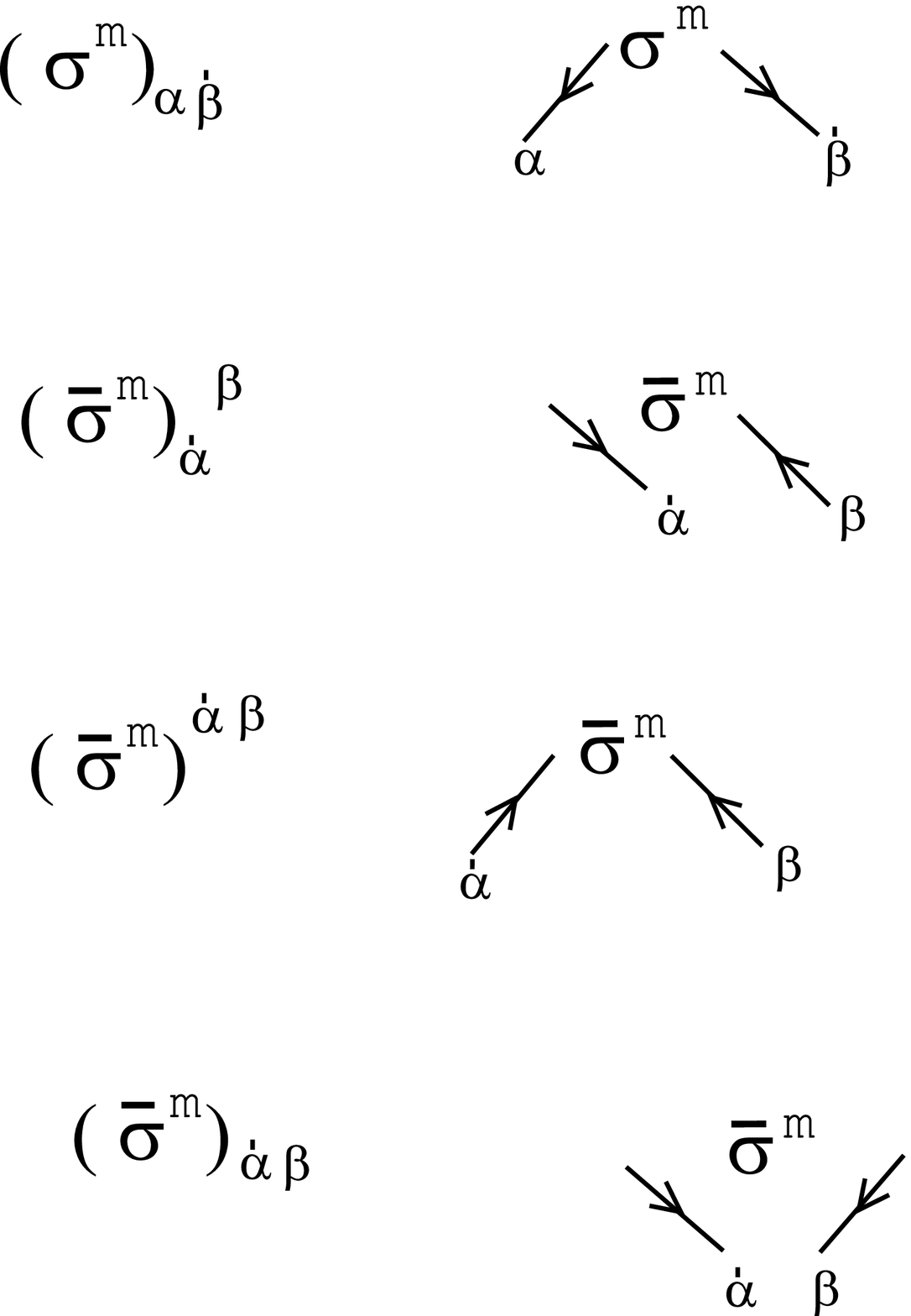}
           }
\label{LF3}
   \caption{
Elements of SL(2,C) $\si$-matrices. 
$(\si^m)_{\al\bedot}$ and $(\sibar^m)^{\aldot\be}$ are
the standard form.
   }
\end{figure}
They are stored as the 2$\times$2 array si[\ ][\ ].\nl
\shortstack[l]{
$\si^m_{~\al\aldot}$\\
si[0,0]=empty\\
si[0,1]=$\al$\\
si[1,0]=empty\\
si[1,1]=$\aldot$\\
siv=m
             }
\q
\shortstack[l]{
$\sibar^{m\aldot\al}$\\
si[0,0]=$\aldot$\\
si[0,1]=empty\\
si[1,0]=$\al$\\
si[1,1]=empty\\
siv=m
             }
\q

Note: The use of even ($\al$) and odd ($\aldot$) integers makes
an important role here. 
The arrangement of spinor suffixes, that is
(left-side suffix, right-side suffix)=(odd, even) or (even, odd), makes
us clear the difference between $\si$ and $\sibar$. 
We should, however, have the relation 
$$\sibar^{m}_{\aldot\al}=\si^{m}_{\al\aldot}$$
 in mind. Hence the above 2quantities
are equivelently expressed as\nl
\shortstack[l]{
$\sibar^m_{~\aldot\al}$\\
si[0,0]=empty\\
si[0,1]=$\aldot$\\
si[1,0]=empty\\
si[1,1]=$\al$\\
siv=m
             }
\q
\shortstack[l]{
$\si^{m\al\aldot}$\\
si[0,0]=$\al$\\
si[0,1]=empty\\
si[1,0]=$\aldot$\\
si[1,1]=empty\\
siv=m
             }
\nl
This ambiguity does not cause any problem because we keep a rule:
\begin{itemize}
\item
In this program, we use only $\si$ (not use $\sibar^m$).\\ 
\item
$\sibar^m$ is used only for the graphical explanation.
\end{itemize}
(4) Superspace coordinate [Symbol: t\ ;\ Dimension: M$^{-1/2}$ ]\nl
The superspace coordinate $\sh^\al$ is exprssed in the same way
as the spinor $\psi^\al$.\nl
\nl
\shortstack[l]{
$\sh^\al$\\
th[0,0]=$\al$\\
th[0,1]=empty\\
th[1,0]=empty\\
th[1,1]=empty
             }  
\q
\shortstack[l]{
$\sh_\al$\\
th[0,0]=empty\\
th[0,1]=$\al$\\
th[1,0]=empty\\
th[1,1]=empty
             }
\q
\shortstack[l]{
$\thbar_\aldot$\\
th[0,0]=empty\\
th[0,1]=empty\\
th[1,0]=empty\\
th[1,1]=$\aldot$
             }
\q
\shortstack[l]{
$\thbar^\aldot$\\
th[0,0]=empty\\
th[0,1]=empty\\
th[1,0]=$\aldot$\\
th[1,1]=empty
             }
\nl
They are graphically expressed by Fig.4.
\begin{figure}[htbp]
\centerline{
\includegraphics*[height=6cm]{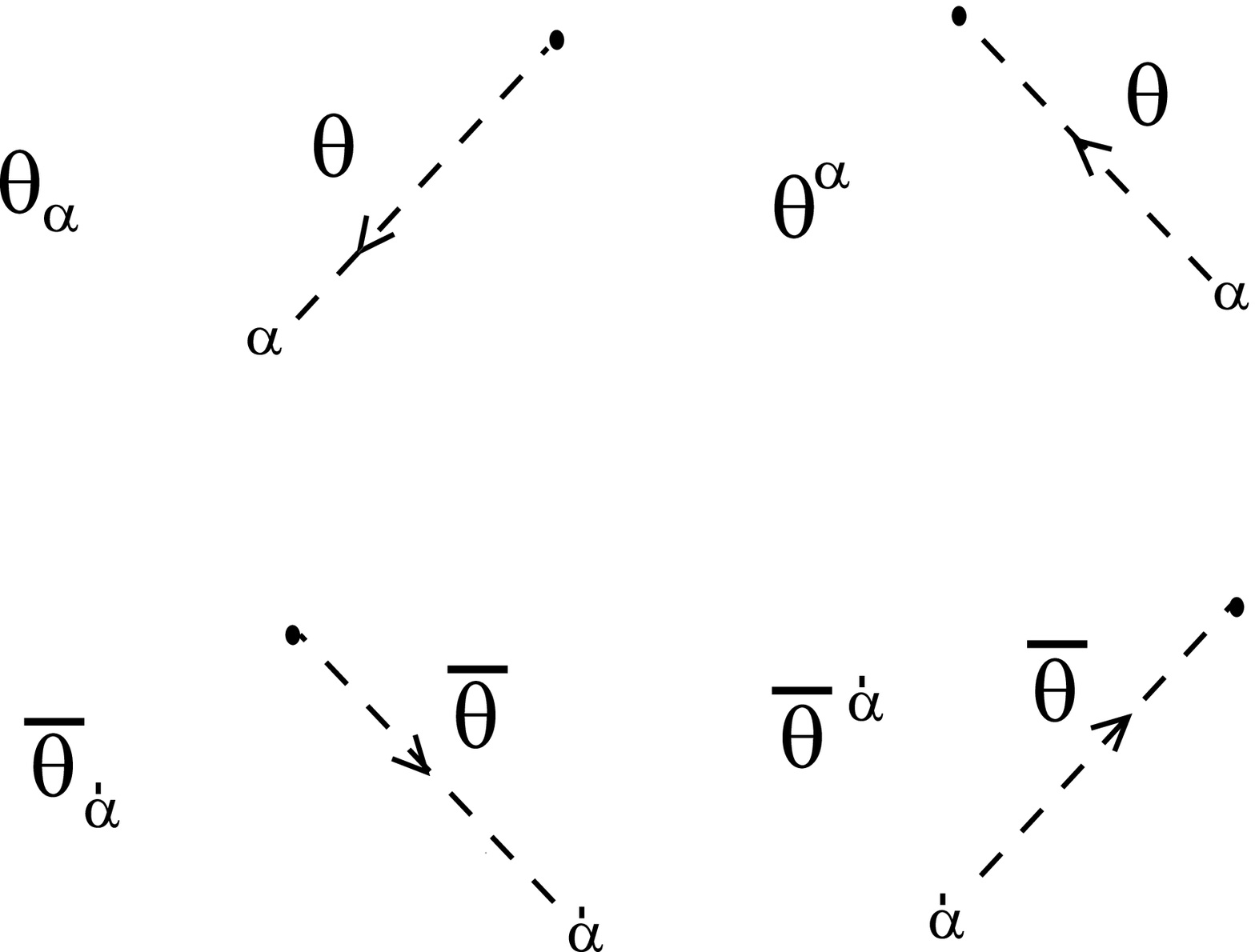}
           }
\label{LF15}
   \caption{
The graphical representation for the
spinor coordinates in the superspace:\ $\sh_\al, \sh^\al, \thbar_\aldot$ and 
$\thbar^\aldot$.
   }
\end{figure}
\nl\nl
(5) Gagino [Symbol: l\ ;\ Dimension: M$^{3/2}$ ]\nl
The photino $\la^\al$ is exprssed in the same way
as the spinor $\psi^\al$.
We take the 2$\times$2 array la[\ ][\ ].\nl
\nl
\shortstack[l]{
$\la^\al$\\
la[0,0]=$\al$\\
la[0,1]=empty\\
la[1,0]=empty\\
la[1,1]=empty
             }  
\q
\shortstack[l]{
$\la_\al$\\
la[0,0]=empty\\
la[0,1]=$\al$\\
la[1,0]=empty\\
la[1,1]=empty
             }
\q
\shortstack[l]{
$\labar_\aldot$\\
la[0,0]=empty\\
la[0,1]=empty\\
la[1,0]=empty\\
la[1,1]=$\aldot$
             }
\q
\shortstack[l]{
$\labar^\aldot$\\
la[0,0]=empty\\
la[0,1]=empty\\
la[1,0]=$\aldot$\\
la[1,1]=empty
             }
\nl\nl
(6) The first derivative of gaugino [Symbol: m\ ;\ Dimension: M$^{5/2}$ ]\nl
The first derivative of the photino is expressed as
the 2$\times$2 array dl[\ ][\ ] and the variable dlv. 
\nl  
\shortstack[l]{
$\pl_m\la^\al$\\
dl[0,0]=$\al$\\
dl[0,1]=empty\\
dl[1,0]=empty\\
dl[1,1]=empty\\
dlv=m
             }
\q
\shortstack[l]{
$\pl_m\la_\al$\\
dl[0,0]=empty\\
dl[0,1]=$\al$\\
dl[1,0]=empty\\
dl[1,1]=empty\\
dlv=m
             }
\q
\shortstack[l]{
$\pl_m\labar_\aldot$\\
dl[0,0]=empty\\
dl[0,1]=empty\\
dl[1,0]=empty\\
dl[1,1]=$\aldot$\\
dlv=m
             }
\q
\shortstack[l]{
$\pl_m\labar^\aldot$\\
dl[0,0]=empty\\
dl[0,1]=empty\\
dl[1,0]=$\aldot$\\
dl[1,1]=empty\\
dlv=m
             }

\vspace{5mm}

{\bf Bosonic Fields}
\nl
(7) Complex scalar [Symbol: A\ ;\ Dimension: M$^1$ ]\nl
The complex scalar field $A$ is expressed 
by one dimensional array A[\ ] with 2 elements.\nl
\shortstack[l]{
A\\
A[0]=1(exist)\\
A[1]=empty
             }  
\q
\shortstack[l]{
A$^*$\\
A[0]=empty\\
A[1]=1(exist)
             }
\nl
where the element-numbers 0,1 correspond to $A$(chiral) or $A^*$(anti-chiral), 
respectively.
\nl\nl
(8) The first derivative of the complex scalar [Symbol: B\ ;\ Dimension: M$^2$ ]\nl
The first derivative of $A$ and $A^*$ are expressed as the one dimensional array
B[\ ] with 2 elements.\nl
\shortstack[l]{
$\pl_m$A\\
B[0]=m\\
B[1]=empty
             }  
\q
\shortstack[l]{
$\pl_m$A$^*$\\
B[0]=empty\\
B[1]=m
             }
\nl\nl
(9) Vector field [Symbol: v\ ;\ Dimension: M$^1$ ]\nl
The vector field(photon) $v^m$ is expressed by one variable v.\nl
\nl
\shortstack[l]{
$v^m$\\
v=m\\
v=empty(non-exist)
             }  
\nl
The lowest expression is taken when the vector field does not appear.
\nl\nl
(10) The first derivative of the vector field [Symbol: w\ ;\ Dimension: M$^2$ ]\nl
The first derivative of $v^m$ is expressed by two variables dv and dvv.\nl
\nl
\shortstack[l]{
$\pl_nv_m$\\
dv=m\\
dvv=n
               }  
\nl\nl
(11) The Dalemberian derivative of A and A$^*$ [Symbol: C\ ;\ Dimension: M$^3$ ]\nl
The Dalemberian derivative of $A$ and $A^*$ are expressed as
the one dimensional array C[\ ] with 2 elements.\nl
\shortstack[l]{
$\pl_m\pl^m$A\\
C[0]=1(exist)\\
C[1]=empty
             }  
\q
\shortstack[l]{
$\pl_m\pl^m$A$^*$\\
C[0]=empty\\
C[1]=1(exist)
             }
\nl\nl
(12) Auxiliary fields [Symbol: F\ ;\ Dimension: M$^2$ ]\nl
The auxiliary fields $F$ and $F^*$ are expressed as\nl
\shortstack[l]{
F\\
F[0]=1(exist)\\
F[1]=empty
             }  
\q
\shortstack[l]{
F$^*$\\
F[0]=empty\\
F[1]=1(exist)
             }
\nl\nl
(13) Real auxilary fields [Symbol: D\ ;\ Dimension: M$^2$ ]\nl
The auxiliary field $D$ (real scalar), which appears in the vector
multiplet, is expressed as\nl
\shortstack[l]{
D\\
D=1(exist)\\
D=empty(non-exist)
             }  \nl

{\bf Term and Component}

Terms in the SUSY calculation are stored as an ordered set of the above quantities. 
The following examples appear in the intermidiate stage of evaluating $W_\al W^\al$
where $W_\al$ is the field strength superfield. (See App.B.)

Example 1

The term $\si^l_{\del\aldot}\si^{m\be\aldot}\si^{s\del}_{~\ \gadot}\si^{n\ \gadot}_{~\be}
\pl_lv_m\pl_sv_n$= 
$-\si^l_{\del\aldot}\sibar^{m\aldot\be}\si^n_{\be\gadot}\sibar^{s\gadot\del}
\pl_lv_m\pl_sv_n$ is stored, in the computer, as follows.\nl
\shortstack[l]{
type[c=0]=s\\
si[c=0,0,1]=1\\
si[c=0,1,1]=2\\
siv[c=0]=51
             }  
\q
\shortstack[l]{
type[c=1]=s\\
si[c=1,0,0]=3\\
si[c=1,1,0]=2\\
siv[c=1]=52
             }  
\q
\shortstack[l]{
type[c=2]=s\\
si[c=2,0,0]=1\\
si[c=2,1,1]=4\\
siv[c=2]=53
             }  
\q
\shortstack[l]{
type[c=3]=s\\
si[c=3,0,1]=3\\
si[c=3,1,0]=4\\
siv[c=3]=54
             }  
\q
\shortstack[l]{
\\
conti-\\
nued\\
below
             }  
\nl
\nl
\shortstack[l]{
type[c=4]=w\\
dv[c=4]=52\\
dvv[c=4]=51
             }  
\q
\shortstack[l]{
type[c=5]=w\\
dv[c=5]=54\\
dvv[c=5]=53
             }  
\nl
($\del\ra 1,\aldot\ra 2,\be\ra 3,\gadot\ra 4;\ l\ra 51,m\ra 52,s\ra 53,n\ra 54$)
\nl
Besides the above ones, the number of components (compno=6) and 
an overall {\it weight} (complex) are necessary to characterize a term. 
One additional coloumn, specified by  the variable c, appears. The variable c 
controls the order of every {\it component}. This {\it ordering}
is important in the calculation involving {\it Grassmannian}
quantities. In the above storing form, 
each component starts with specifying the type: s, w, $\cdots$. 
The one dimensional array type[\ ] is used for the purpose. 
This term graphically appears in the output (before further reduction) 
as shown in Fig.5.

\begin{figure}[htbp]
\centerline{
\includegraphics*[height=3cm]{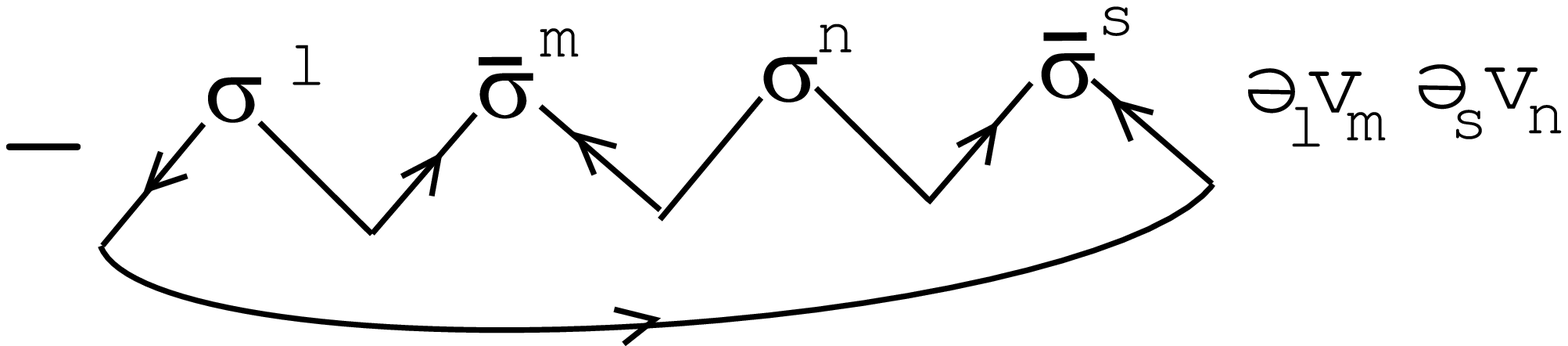}
             }
\label{LssbssbR}
   \caption{
The graphical expression of
$-\si^l_{\del\aldot}\sibar^{m\aldot\be}\si^n_{\be\gadot}\sibar^{s\gadot\del}
\pl_lv_m\pl_sv_n$. 
   }
\end{figure}

In Fig.5
, we see all dummy suffixes disappear and the advantage of the graphical
expression is manifest. The chirality can be read from the shape of the 
directed-line graph.

Example 2

The term 
$\si^{l\del}_{\ \aldot}\si^{m~\aldot}_{\ \del}D\pl_lv_m$=
$-\si^l_{\del\aldot}\sibar^{m\aldot\del}D
\pl_lv_m$ is stored, in the computer, as follows.\nl
\shortstack[l]{
type[c=0]=s\\
si[c=0,0,0]=1\\
si[c=0,1,1]=4\\
siv[c=0]=53
             }  
\q
\shortstack[l]{
type[c=1]=s\\
si[c=1,0,1]=1\\
si[c=1,1,0]=4\\
siv[c=1]=54
             }  
\q
\shortstack[l]{
type[c=2]=D\\
D[c=2]=1
             }  
\q
\shortstack[l]{
type[c=3]=w\\
dv[c=3]=54\\
dvv[c=3]=53
             }  
\nl
($\del\ra 1,\aldot\ra 4;\ l\ra 53,m\ra 54$)\nl
In this case, the number of components is 4 (compno=4) and there 
is a weight. This one graphically appears in the output (before further reduction) 
as shown in Fig.6.
\begin{figure}[htbp]
\centerline{
\includegraphics*[height=2cm]{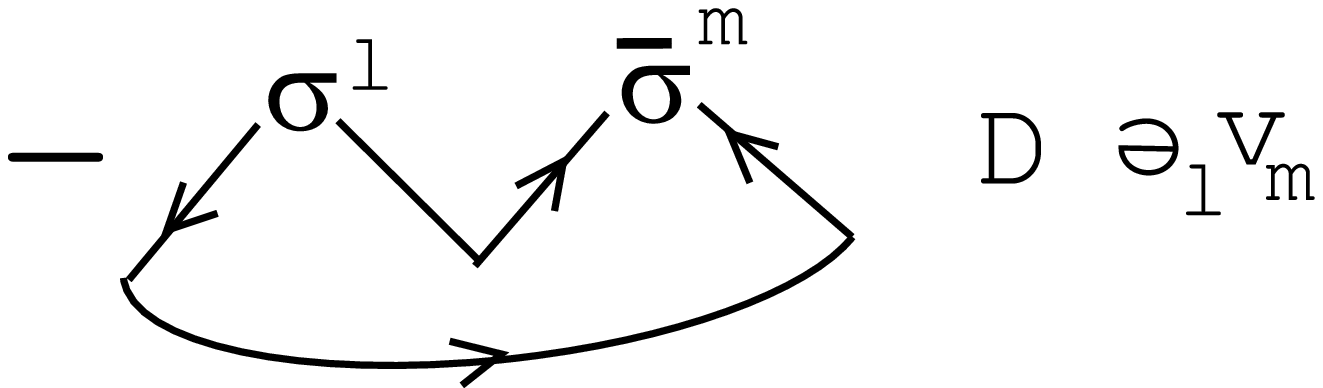}
           }
\label{LssbR}
   \caption{
The graphical expression of
$-\si^l_{\del\aldot}\sibar^{m\aldot\del}D
\pl_lv_m$. 
   }
\end{figure}

These two examples clearly show the advantage of the graph representation
over the conventional one using indicies. The new representation
discriminates each term not by the suffixes but by the shape of the graph.
It means the importance of indices of the graph such as the number
of the chiral-loop. (In the above examples, the number is 1.)

\section{Field name, id-number and Grassmannian property}\label{sec:grass}
One advantage of C-language is the efficient  
manipulation of characters. 
The control of characters is important for specifying various
operators (fields) appearing in SUSY theories. 
As described previously, we assign every operator a symbol which
is expressed by a character. The appearance of many fields, which is
a big cause complicating SUSY theories, can be neatly handled by the use of the
character name. We also introduce the id-number for every operator. 
(See the list below) 

For every operator, we assign the Grassmann number:\ +1(commuting) or -1(anti-commuting). 
This assignment is exploited in the process of {\it moving a component}
, within a term, {\it respecting the Grassmannian property of operators}.

We list the assignment in TABLE 1 with the dimension of operators.
 
\begin{table}
\begin{center}
\begin{tabular}{ccc}
    Fid[0]='t';   &   DIM[0]=0.0;     &      Grassmann[0]=-1; \\
    Fid[1]='s';   &   DIM[1]=0.0;     &      Grassmann[1]=1;  \\
    /* <Chiral */ &   & \\
    Fid[100]='p'; &   DIM[100]=(float)3/2; & Grassmann[100]=-1;\\
    Fid[101]='q'; &   DIM[101]=(float)5/2; & Grassmann[101]=-1;\\
    Fid[102]='A'; &   DIM[102]=1.0;   &      Grassmann[102]=1;  \\
    Fid[103]='B'; &   DIM[103]=2.0;   &      Grassmann[103]=1;  \\
    Fid[104]='C'; &   DIM[104]=3.0;   &      Grassmann[104]=1;  \\
    Fid[105]='F'; &   DIM[105]=2.0;   &      Grassmann[105]=1;  \\
    /* <Chiral2 */ & & \\
    Fid[120]='P'; &   DIM[120]=(float)3/2; & Grassmann[120]=-1; \\
    Fid[121]='Q'; &   DIM[121]=(float)5/2; & Grassmann[121]=-1;  \\
    Fid[122]='a'; &   DIM[122]=1.0;    &     Grassmann[122]=1;  \\
    Fid[123]='b'; &   DIM[123]=2.0;    &     Grassmann[123]=1;  \\
    Fid[124]='c'; &   DIM[124]=3.0;    &     Grassmann[124]=1;  \\
    Fid[125]='f'; &   DIM[125]=2.0;    &     Grassmann[125]=1;  \\
    /* <Vector */ & & \\
    Fid[140]='l'; &   DIM[140]=(float)3/2; & Grassmann[140]=-1; \\
    Fid[141]='m'; &   DIM[141]=(float)5/2; & Grassmann[141]=-1; \\
    Fid[142]='D'; &   DIM[142]=2.0;        & Grassmann[142]=1;  \\
    Fid[143]='v'; &   DIM[143]=1.0;        & Grassmann[143]=1;  \\
    Fid[144]='w'; &   DIM[144]=2.0;        & Grassmann[144]=1;  \\
 & & \\
\multicolumn{3}{c}{TABLE 1\ Definition of FieldID, Dimension, Grassmann No}
\end{tabular}
\end{center}
\end{table}

\section{Graph Indices: vpairno, NcpairO, NcpairE, closed-chiral-loop-No, GrNum}\label{sec:gind}

SigmaContraction (B),(C),(D) and (E)

In the process of SUSY calculation, there appear graphs connected by
directed lines (chiral suffixes contraction) and by (non-directed) dotted lines
(vector suffixes contraction). We can classify them by some {\it graph indices}. 
\begin{description}
\item[vpairno]The number of  vector-suffix contractions.
\item[NcpairO]The number of chiral-suffix contractions. This is equal to the number
of left-directed wedges.
\item[NcpairE]The number of anti-chiral-suffix contractions. This is equal to the number
of the right-directed wedges.
\item[closed-chiral-loop-No]The closed-chiral-loop is the case that the directed lines 
, connected by $\si$ or $\sibar$, make a loop. In this case NcpairO=NcpairE. 
The number of closed chiral loops is defined to this index. 
\item[GrNum]A group is defined to be a set of $\si$'s or $\sibar$'s which
are connected by directed lines. The number of groups is defined to be GrNum.
\end{description} 

In TABLE 2-4, we list the classification of the product of $\si$'s using
the graph indices defined above. 

\begin{table}
\begin{center}
\begin{tabular}{c|c|c|c}
\hline
vpairno & NcpairO & NcpairE & figure\\
\hline
      &   0     &    0     &  $\graph{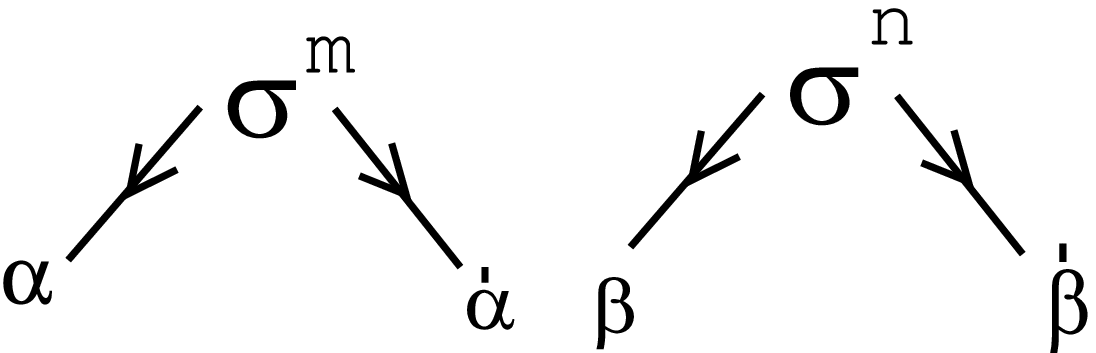}$  \\
\cline{2-4}
  0   &   0      &   1     &  $\graph{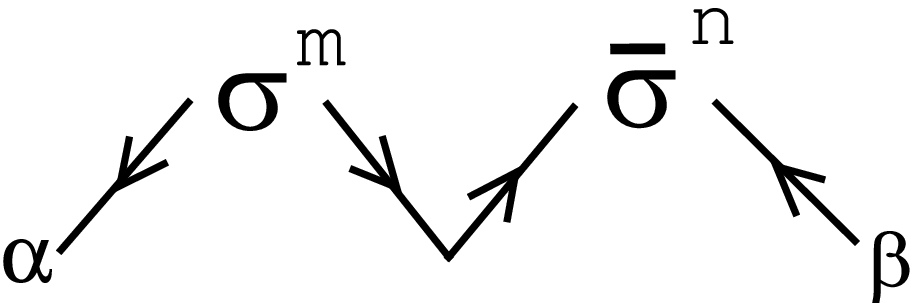}$  \\
\cline{2-4}
       &   1     &   0      &  $\graph{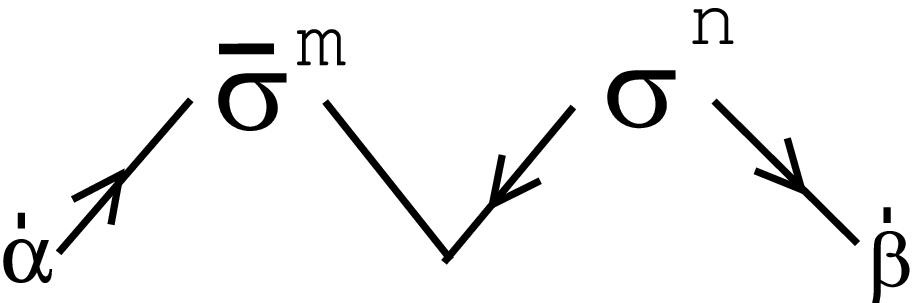}$  \\
\cline{2-4}
       &   1      &   1     &  $\graph{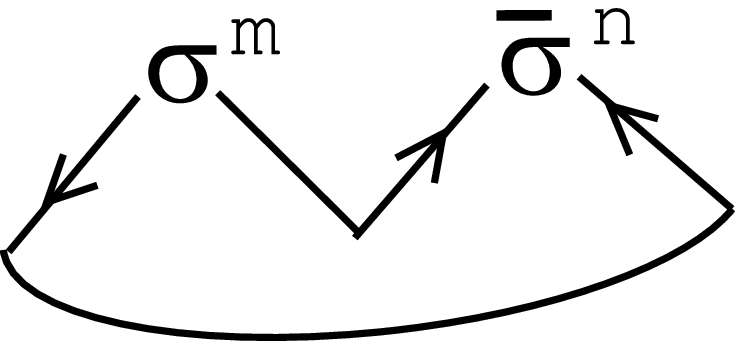}\ =-2\eta^{mn}$  \\
\hline
       &   0       &  0      &  $\graph{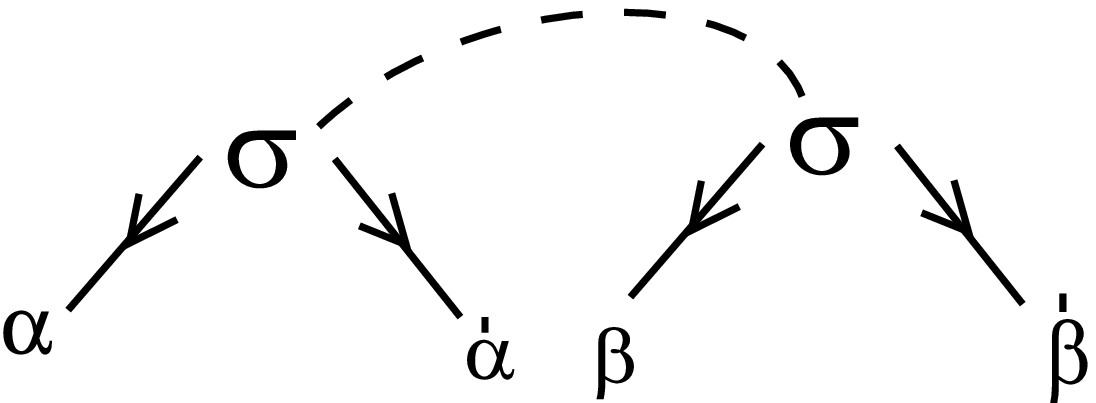}\ =-2\ep_{\al\be}\ep_{\aldot\bedot}$  \\
\cline{2-4}
   1   &   0      &    1     &  $\graph{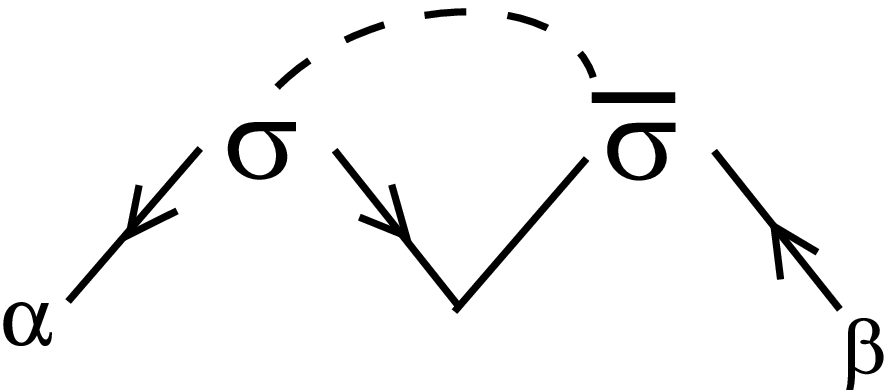}\ =-4\del^\be_\al$  \\
\cline{2-4}
       &    1      &    0     &  $\graph{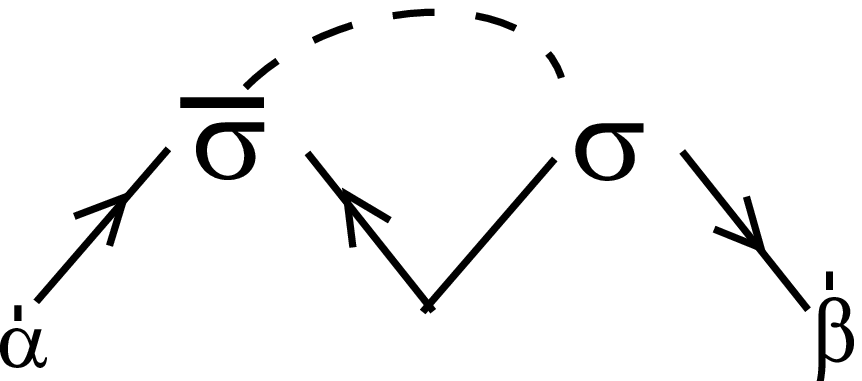}\ =-4\del_\bedot^\aldot$  \\
\cline{2-4}
        &    1     &    1     &   $\graph{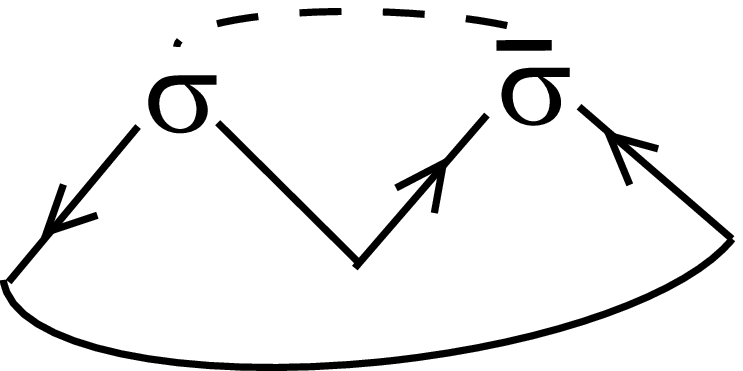}\ =-8$  \\
\hline
\multicolumn{4}{c}{}\\
\multicolumn{4}{c}{TABLE 2\ Classification of the product of 2 sigma matrices (nsi=2).
                  }
\end{tabular}
\end{center}
\end{table}

\begin{table}
\begin{center}
\begin{tabular}{c|c|c|c}
\hline
vpairno & NcpairO & NcpairE & figure\\
\hline
      &   0     &    0     &  $\graph{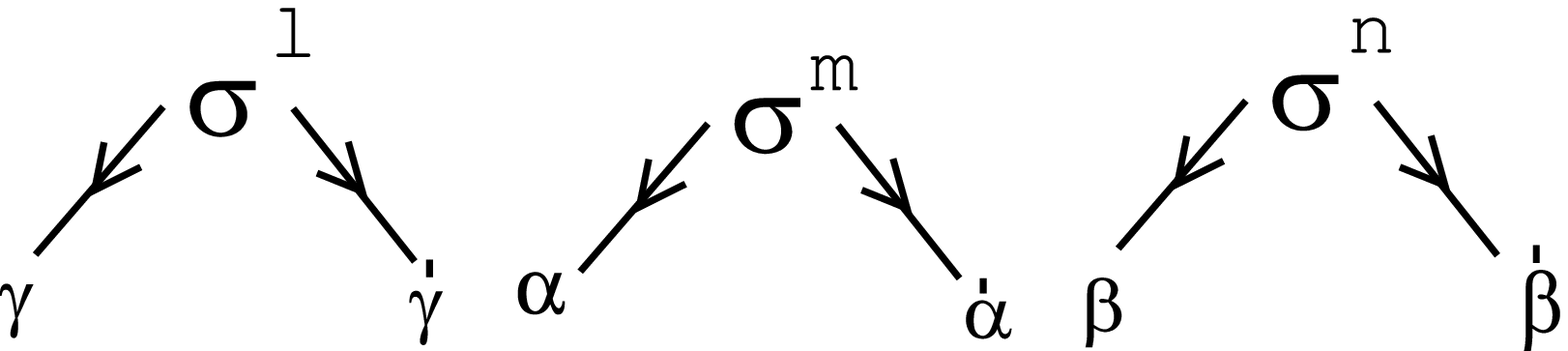}$  \\
\cline{2-4}
  0   &   0      &   1     &  $\graph{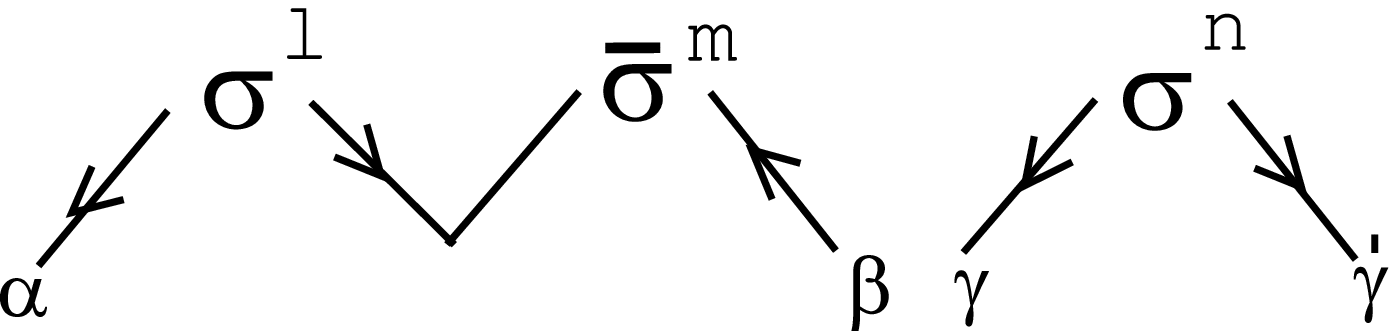}$  \\
\cline{2-4}
       &   1     &   0      &  $\graph{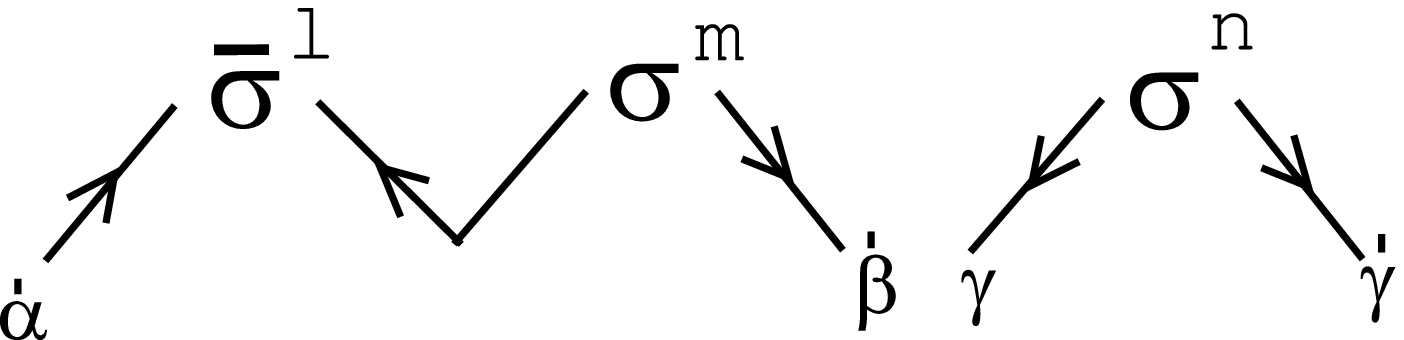}$  \\
\cline{2-4}
       &   1      &   1     &  
\shortstack[l]{
closed-chiral-loop No =1\\
$\graph{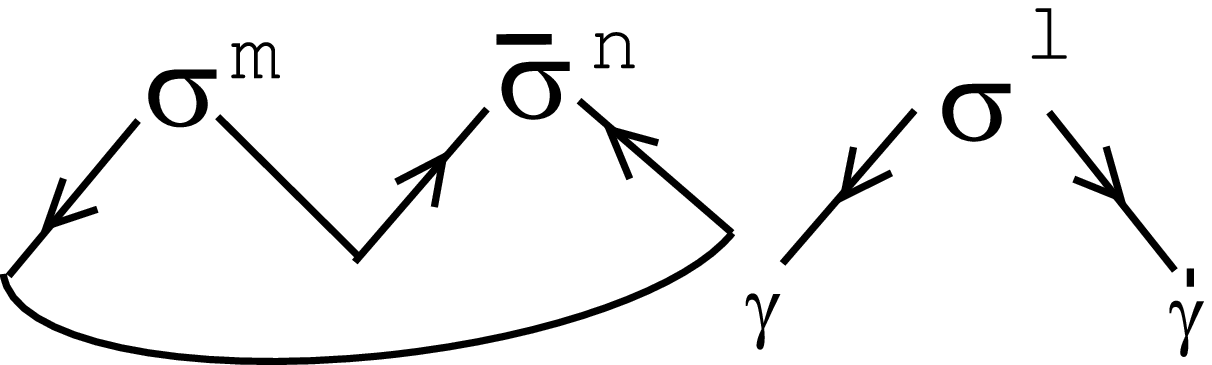}$  }
                                  \\
\cline{4-4}
       &          &         &   
\shortstack[l]{
closed-chiral-loop No =0\\
$\graph{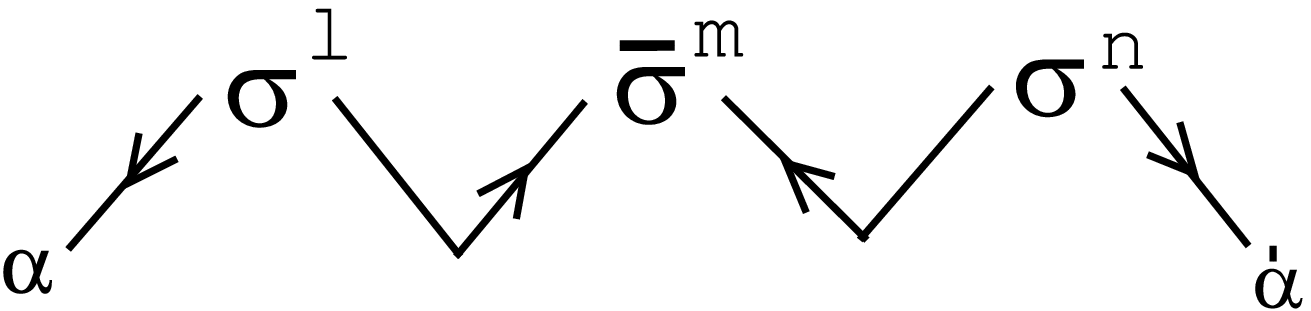}$  }
                                  \\
\hline
       &   0       &  0      &  $-2\ep_{\al\be}\ep_{\aldot\bedot}\ \graph{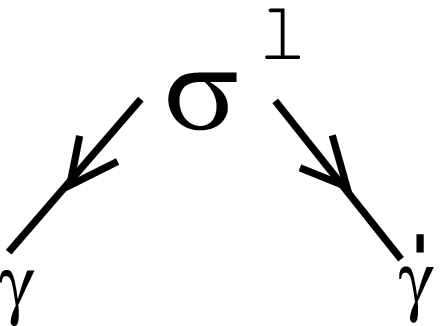}$  \\
\cline{2-4}
   1   &   0      &    1     &  $-4\del^\be_\al\ \graph{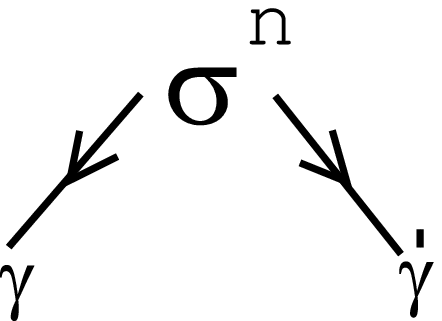}$  \\
\cline{2-4}
       &    1      &    0     &  $-4\del^\aldot_\bedot\ \graph{sig3Dr}$  \\
\cline{2-4}
        &    1     &    1     &   $-8\ \graph{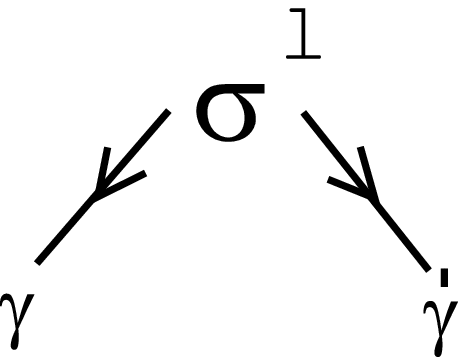}$  \\
\hline
\multicolumn{4}{c}{}\\
\multicolumn{4}{c}{TABLE 3\ Classification of the product of 3 sigma matrices (nsi=3).}
\end{tabular}
\end{center}
\end{table}

\begin{table}
\begin{center}
\begin{tabular}{c|c|c|c}
\hline
vpairno & NcpairO & NcpairE & figure\\
\hline
      &   0     &    0     &  $\graph{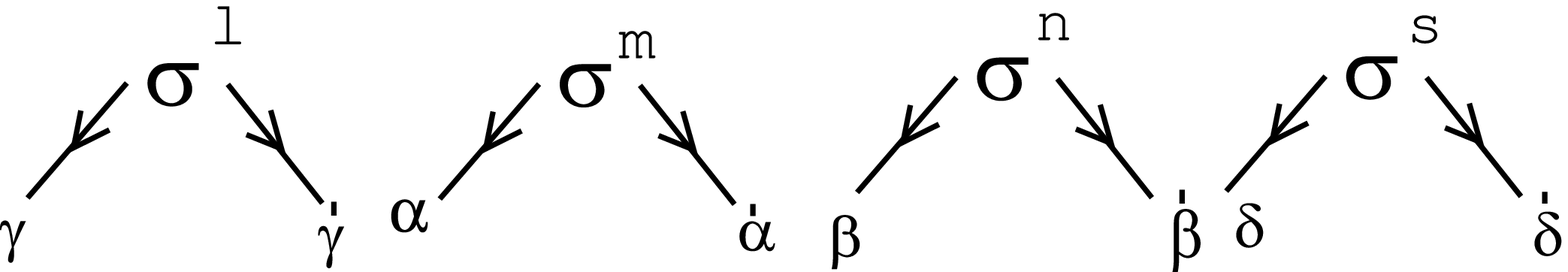}$  \\
\cline{2-4}
      &   0      &   1     &  $\graph{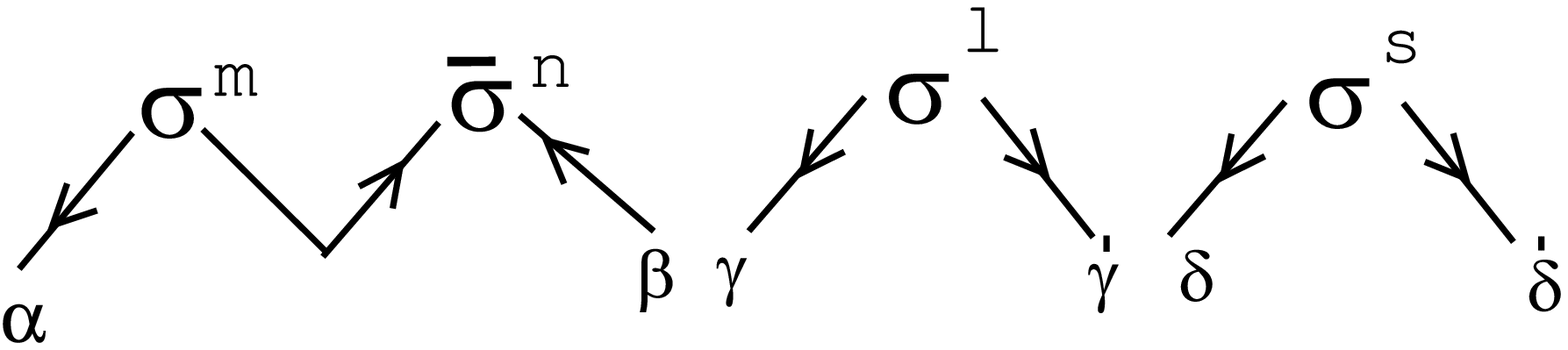}$  \\
\cline{2-4}
      &   1      &   0     &  $\graph{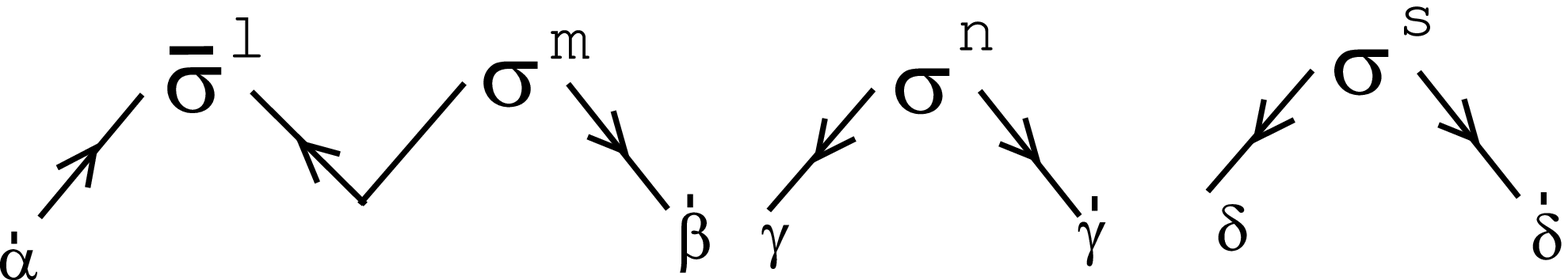}$  \\
\cline{2-4}
      &          &         &  
\shortstack[l]{
GrNum=2, Division=(2,2)\\
$\graph{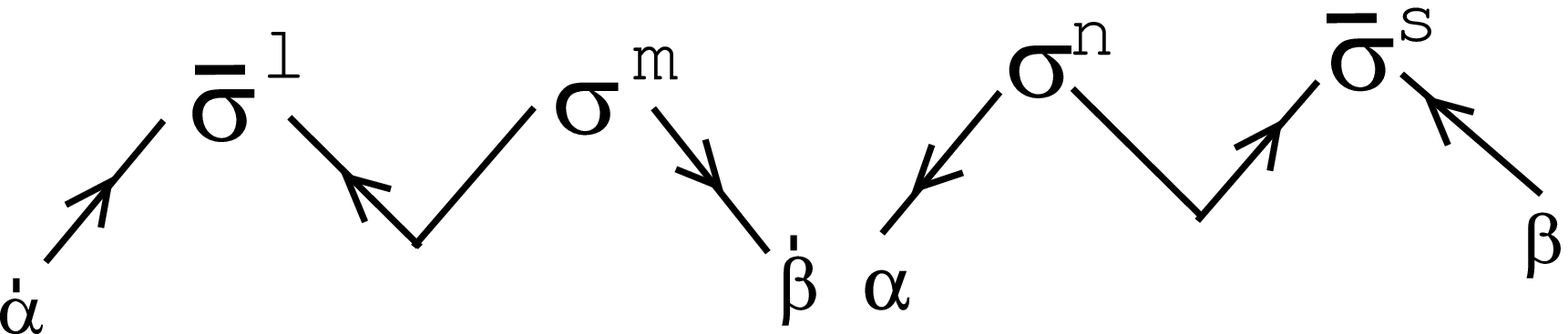}$ } 
                                                    \\
\cline{4-4}
      &   1      &   1     &  
\shortstack[l]{
GrNum=2, Division=(3,1)\\
$\graph{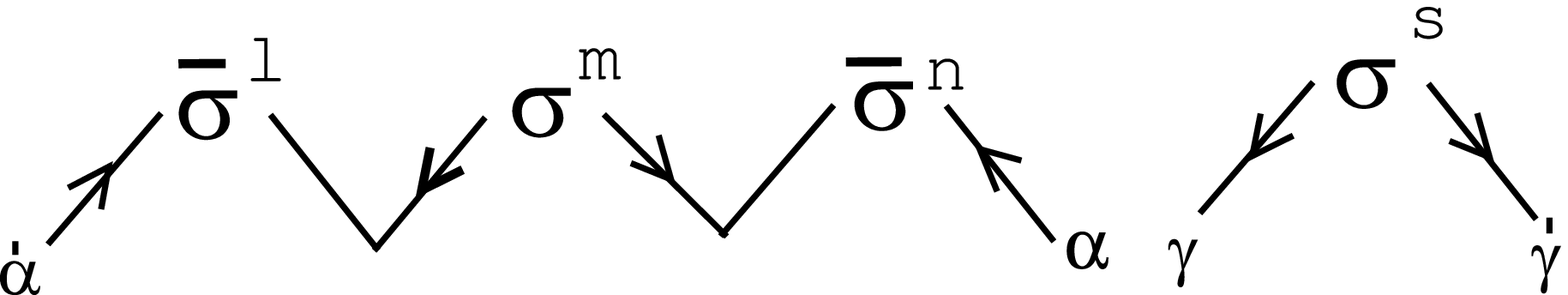}$}                                   \\
\cline{4-4}
      &        &       &  
\shortstack[l]{
GrNum=3, Division=(2,1,1)\\
$\graph{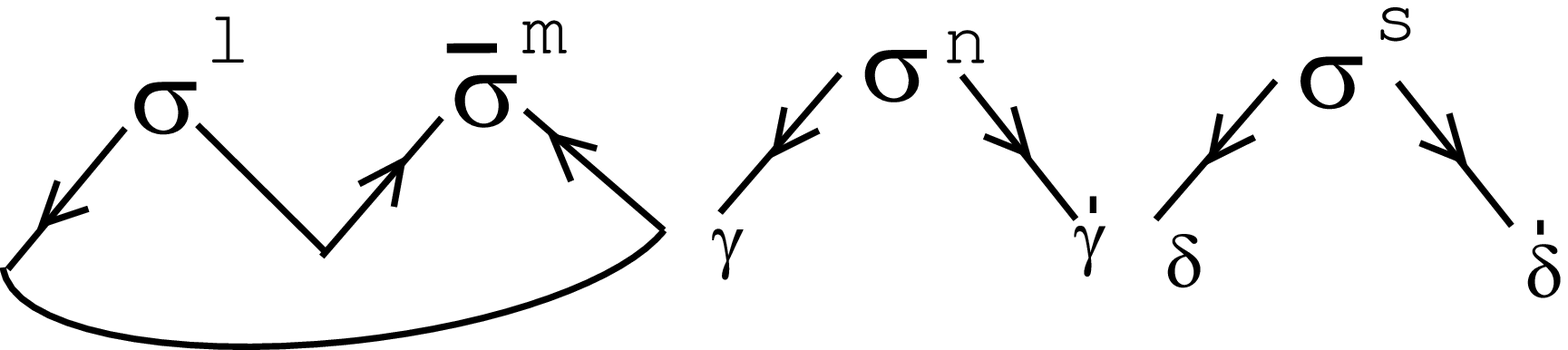}$ }                               \\
\cline{2-4}
  0   &   2      &   0     &  $\graph{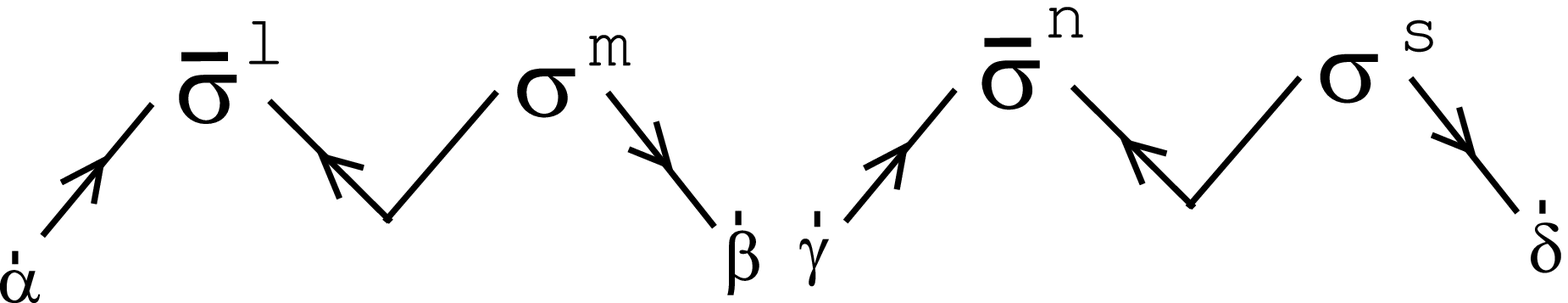}$  \\
\cline{2-4}
       &   0     &   2      &  $\graph{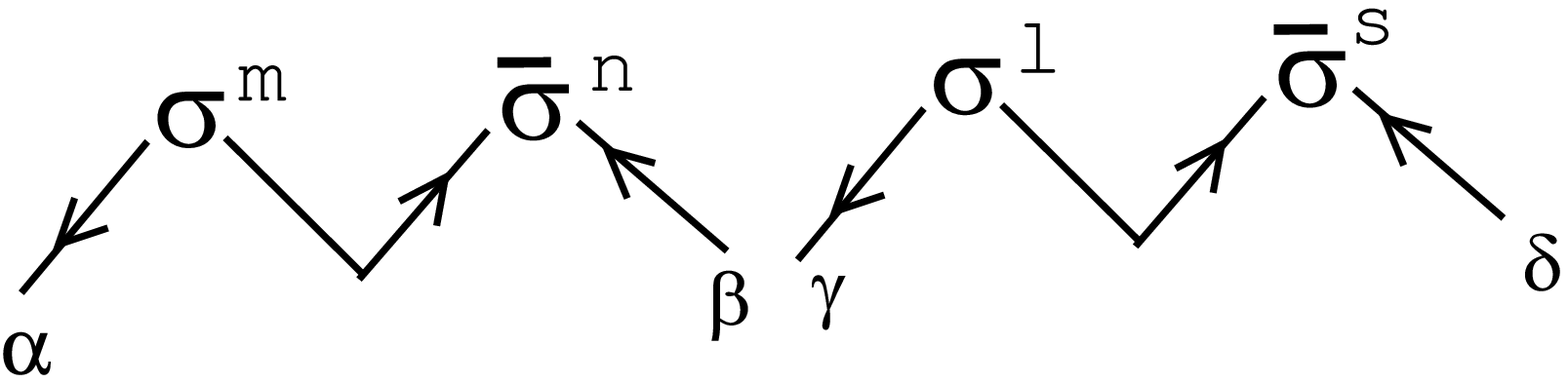}$  \\
\cline{2-4}
       &   1      &   2     &   GrNum=1, $\graph{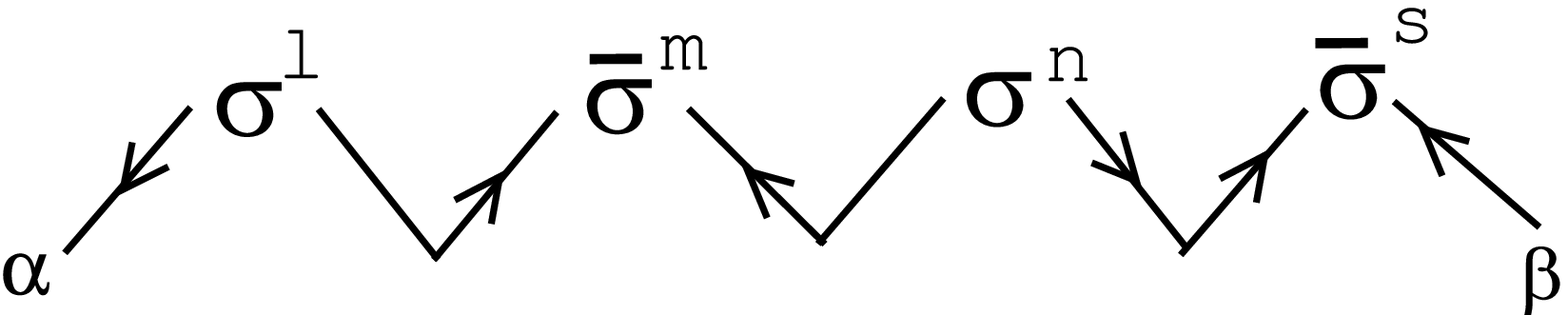}$  \\
\cline{4-4}
       &         &        &  GrNum=2, $\graph{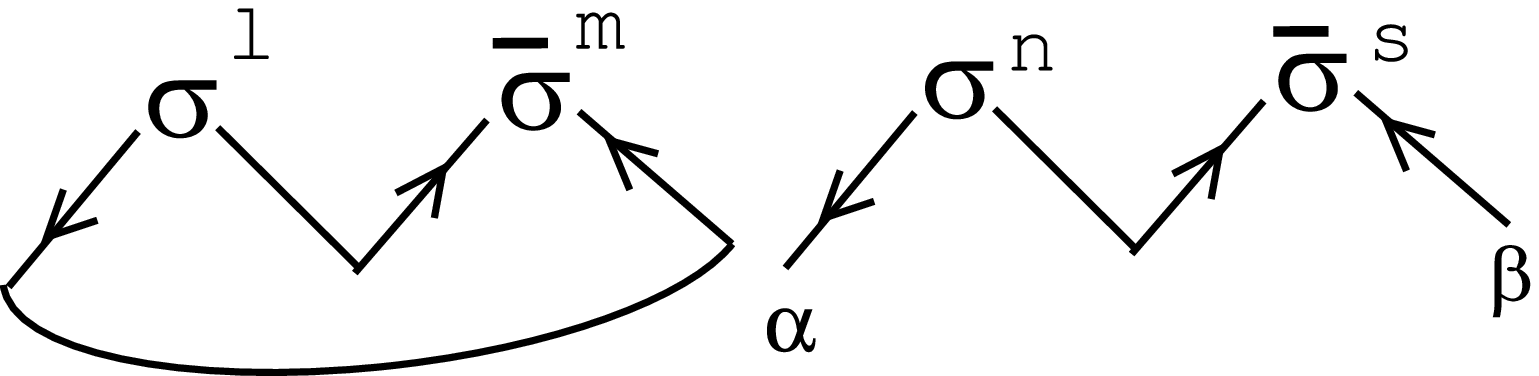}$     \\
\cline{2-4}
       &    2    &     1   &  GrNum=1, $\graph{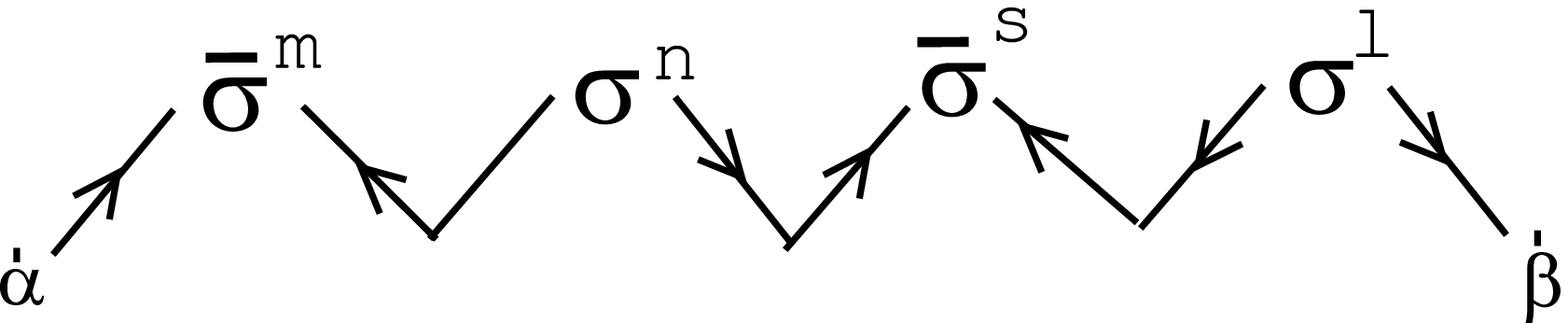}$  
                                                   \\
\cline{4-4}
       &          &         &   GrNum=2, $\graph{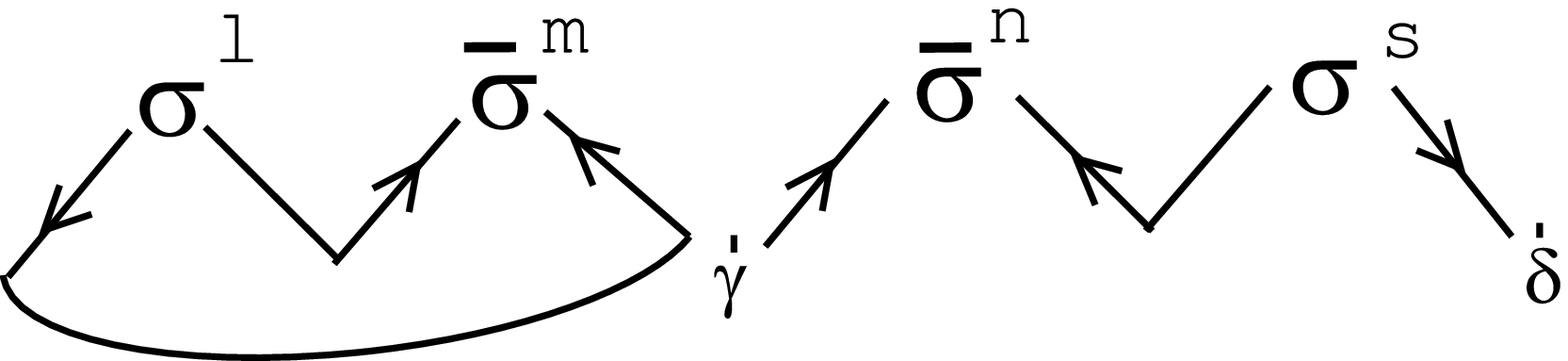}$  
                                                     \\
\cline{2-4}
       &    2     &   2     &   GrNum=1, $\graph{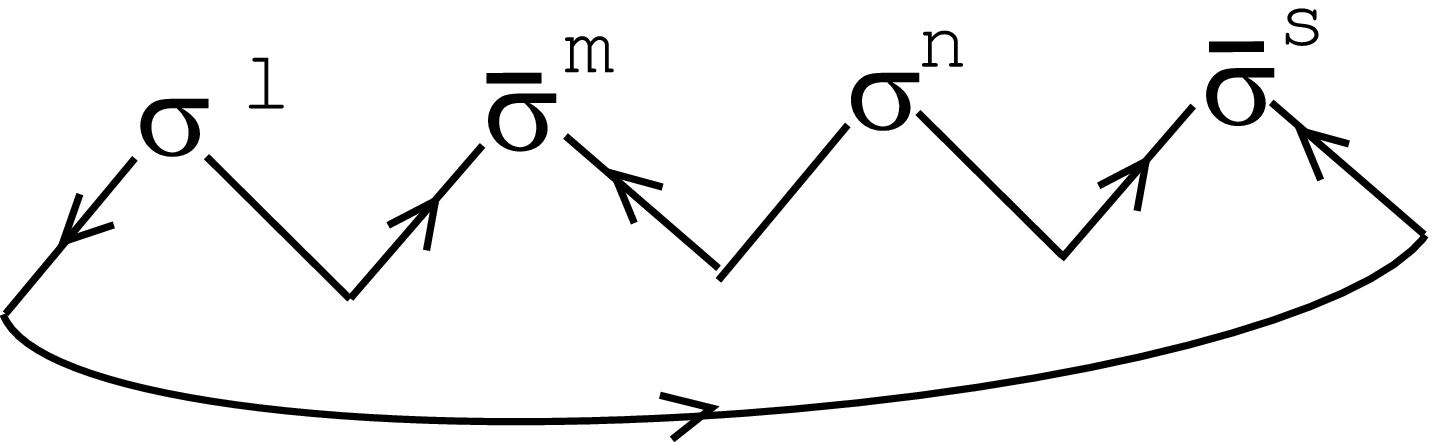}$  
                                                   \\
\cline{4-4}
       &          &         &   GrNum=2,\ $\graph{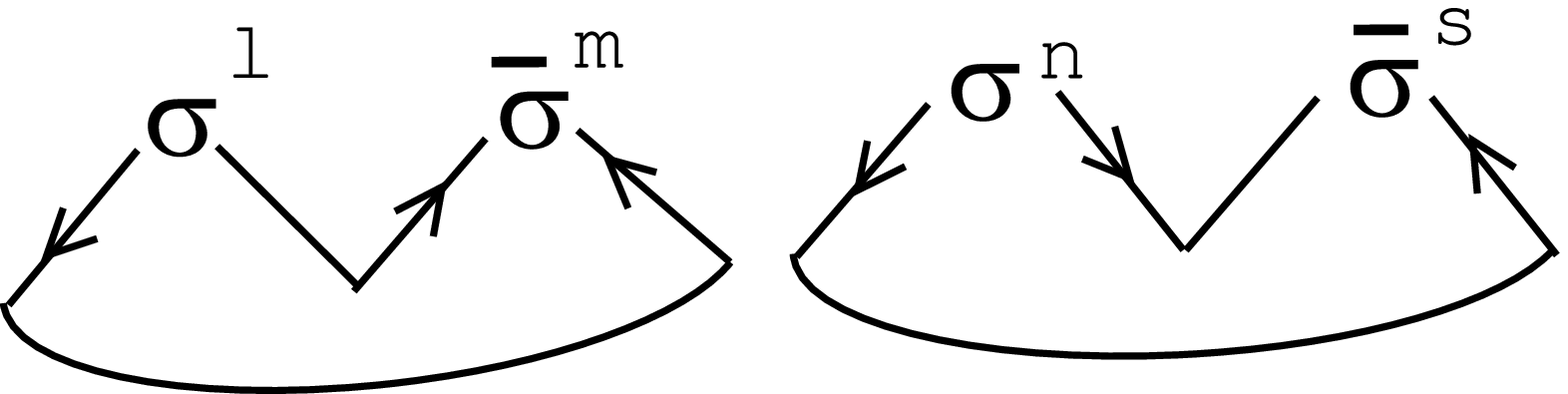}$  
                                                     \\
\hline
\multicolumn{4}{c}{}\\
\multicolumn{4}{c}{TABLE 4\ Classification of the product of 4 sigma matrices with}\\
\multicolumn{4}{c}{\q\q no vector-suffix contraction (nsi=4, vpairno=0). }
\end{tabular}
\end{center}
\end{table}

These tables clearly show the $\si$-matrices play an important role
to connect the chiral world and the space-time (Lorentz) world.

\section{Superspace coordinates}\label{sec:scoordinate}
Supersymmetry is most manifestly expressed in the superspace 
$(x^m,\sh,\thbar)$. $\sh_\al=\ep_\ab\sh^\be$, 
$\thbar^\aldot=\ep^{\aldot\bedot}\thbar_\bedot$ are spinorial
coordinates. 
They satisfy the relations graphically shown in Fig.7.
\begin{figure}[htbp]
\centerline{
\includegraphics*[height=8cm]{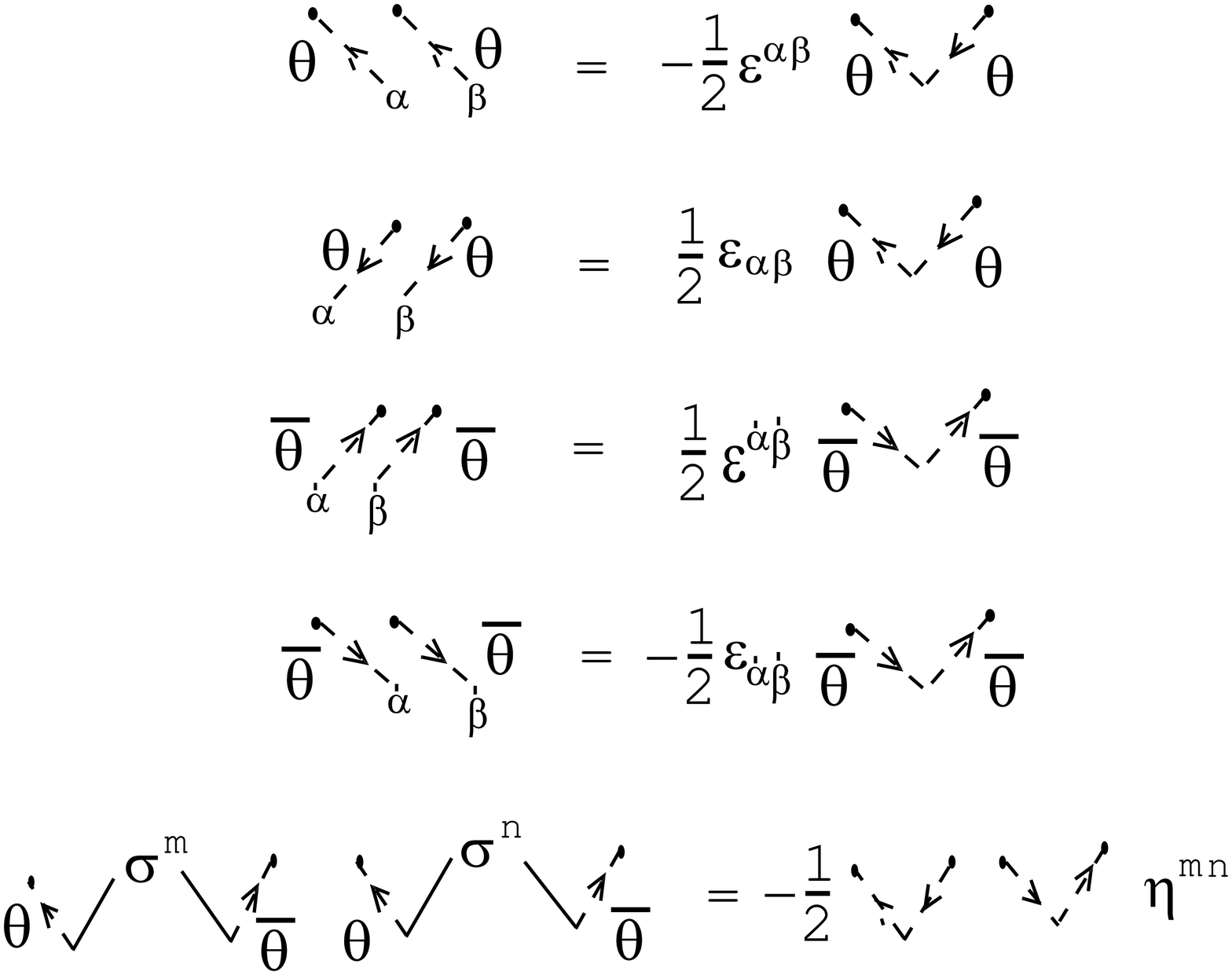}
           }
\label{LF16}
   \caption{
The graphical rules for the spinor coordinates:\ 
$\sh^\al\sh^\be=-\half\ep^\ab\sh\sh,~\sh_\al\sh_\be=\half\ep_\ab\sh\sh,~
\thbar^\aldot\thbar^\bedot=\half\ep^{\aldot\bedot}\thbar\thbar,~
\thbar_\aldot\thbar_\bedot=-\half\ep_{\aldot\bedot}\thbar\thbar,~
\sh\si^m\thbar\sh\si^n\thbar=-\half\sh\sh\thbar\thbar\eta^{mn}$.
   }
\end{figure}
These relations are exploited in the program in order to 
sort the SUSY quantities with respect to the power of $\sh\sh$ and $\thbar\thbar$. 
( For further detail, see the subsection thBthBthth of App.A. )
\section{Treatment of Metrics: $\ep_{\al\be}$, 
$\ep_\al^{~\be}=\ep^\be_{~\al}=\del_\al^\be$, $\eta^{mn}$
and the totally anti-symmetric tensor $\ep^{lmns}$}
For the totally anti-symmetric tensor $\ep^{lmns}$, we introduce
one dimensional array ep[\ ] with 4 components.

\shortstack[l]{
$\mbox{\q\q\q}$\\
$\mbox{\q\q\q}$\\
$\mbox{\q\q\q}$\\
$\mbox{\q\q\q}$
             }  
\q
\shortstack[l]{
ep[0]=l\\
ep[1]=m\\
ep[2]=n\\
ep[3]=s
             }
\q
\shortstack[l]{
$\mbox{\q\q\q}$\\
$\mbox{\q\q\q}$\\
$\mbox{\q\q\q}$\\
$\mbox{\q\q\q}$
             }  
\q
\shortstack[l]{
\\
Symbol: \ e\\
\\
$\mbox{\q\q\q}$
             }
\nl
\nl
This term appears in Sec.\ref{sec:sigma} and produces 
topologically important terms such as
$v_{lm}{\tilde v}^{lm}=\ep^{lmns}v_{lm}v_{ns}$.

As for the metric of the chiral suffix, we do {\it not} introduce specific arrays. They play a role of
raising or lowering suffixes, which can be encoded in the upper (0)
and lower (1) code in arrays. For the Lorentz metric $\eta^{mn}$, we do not
need to much care for the discrimination between the upper and lower suffixes because
of the even-symmetry with respect to the change of the Lorentz suffixes($\eta^{mn}=\eta^{nm}$).

\section{Sigma matrices}\label{sec:sigma}

Let us express important relations valid between products of
sigma matrices graphically. 
In Fig.8
, the symmetric combination of $\sibar^m\si^n$
are shown as the basic spinor algebra.
\begin{figure}
\centerline{ 
\includegraphics*[height=4cm]{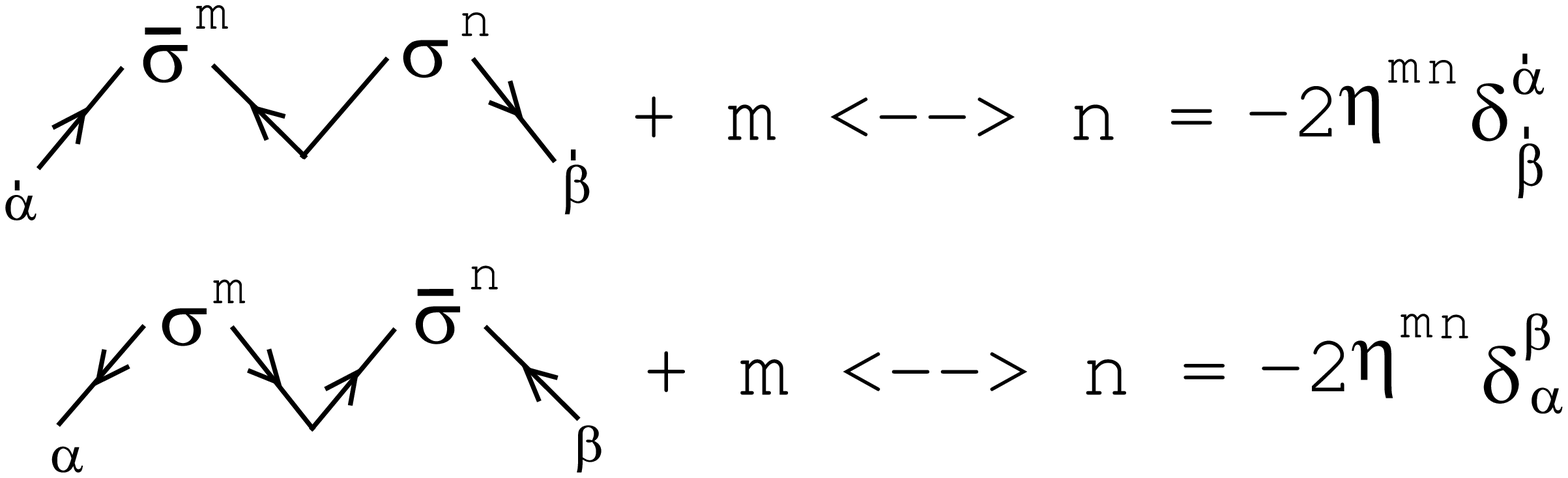}
           }
\label{LF10}
   \caption{
A graphical formula of the basic spinor algebra. 
$(\sibar^m)^{\aldot\be}(\si^n)_{\be\bedot}+m\change n
=-2\eta^{mn}\del^\aldot_\bedot$, and 
$(\si^m)_{\al\aldot}(\sibar^n)^{\aldot\be}+m\change n
=-2\eta^{mn}\del^\be_\al$. 
   }
\end{figure}
The antisymmetric combination gives the generators
of the Lorentz group, $\si^{nm},\sibar^{nm}$.
\begin{eqnarray}
(\si^{nm})_\al^{~\be}=\fourth\left\{
\graph{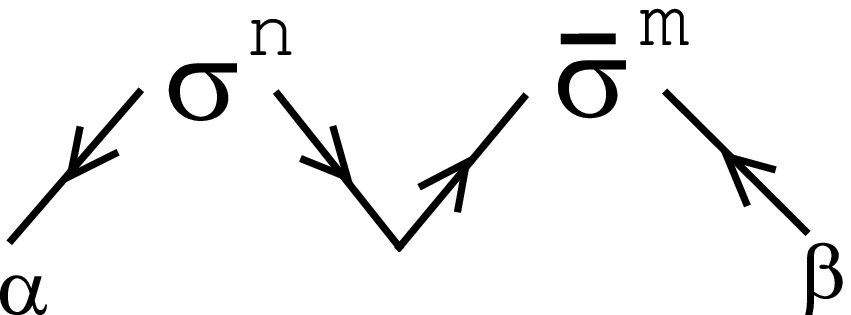}-m\change n\right\}\com\nn
(\sibar^{nm})^\aldot_{~\bedot}=
\fourth\left\{
\graph{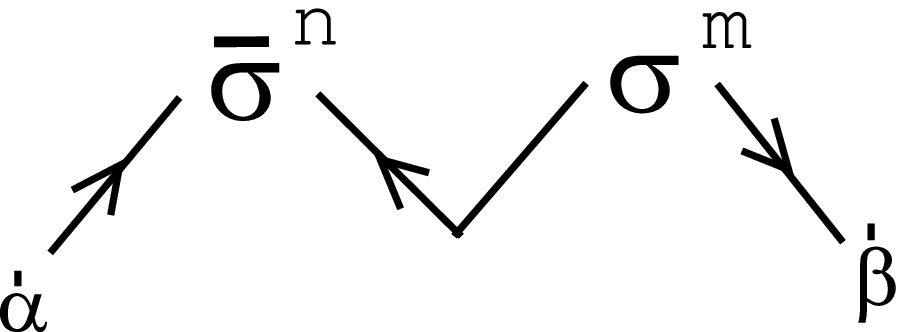}-m\change n\right\}\com
\label{def2}
\end{eqnarray}

The "reduction" formulae (from the cubic $\si'$s to 
the linear one) are expressed as in Fig.9.
\begin{figure}
\centerline{ 
\includegraphics*[height=8cm]{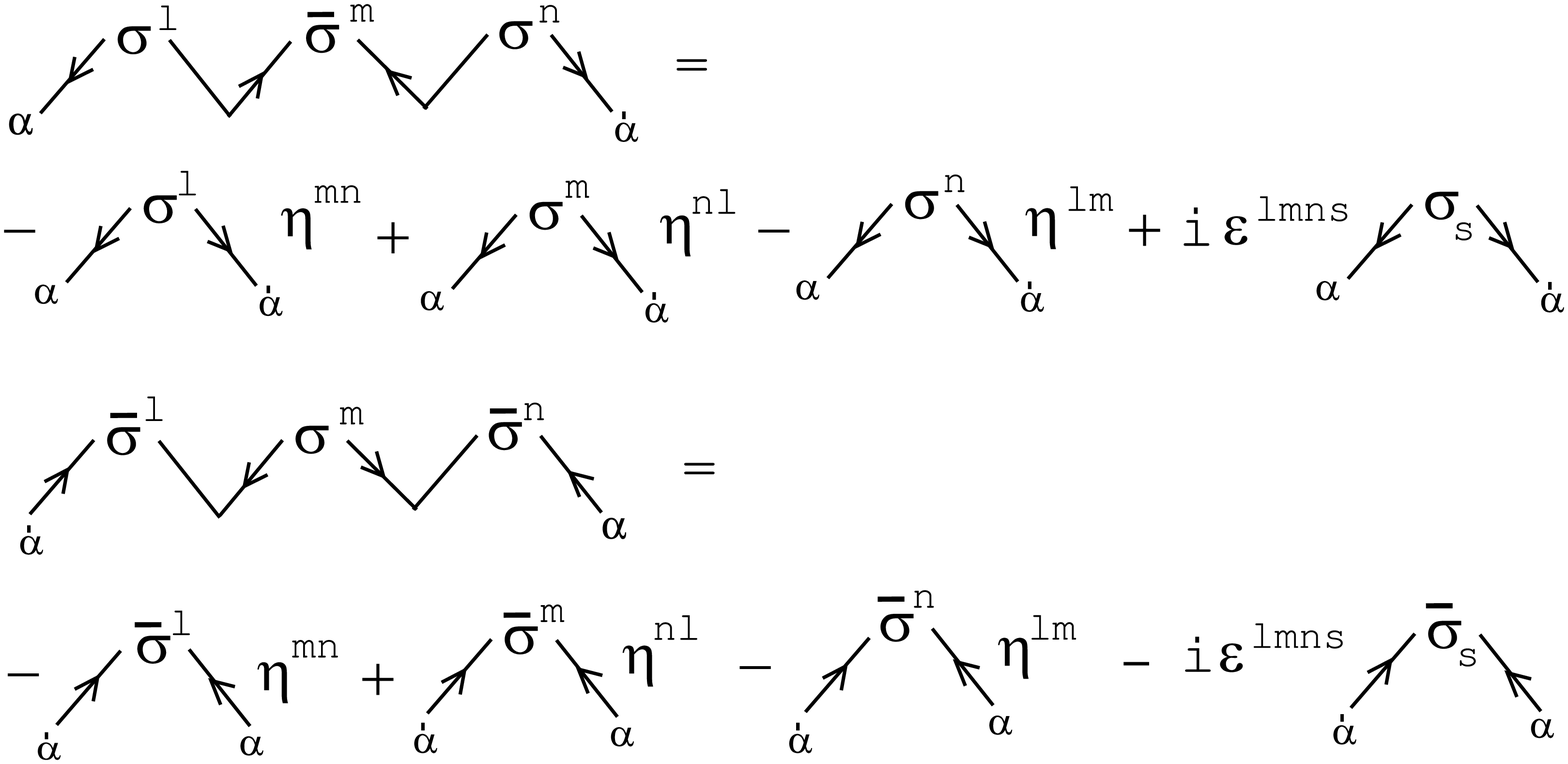}
            }
\label{LF12R}
   \caption{
Two relations:\ 
1) $\si^l\sibar^m\si^n=-\si^l\eta^{mn}+\si^m\eta^{nl}
-\si^n\eta^{lm}+i\ep^{lmns}\si_s$,\ 
2) $\sibar^l\si^m\sibar^n=-\sibar^l\eta^{mn}+\sibar^m\eta^{nl}
-\sibar^n\eta^{lm}-i\ep^{lmns}\sibar_s$.
   }
\end{figure}
From Fig.9
, we notice any chain of $\si'$s
can always be expressed by less than three $\si'$s. 
The appearance of the 4th rank anti-symmetric tensor $\ep^{lmns}$
is quite illuminating. 
The {\it completeness} relations are expressed as in Fig.10.
\begin{figure}
\centerline{ 
\includegraphics*[height=4cm]{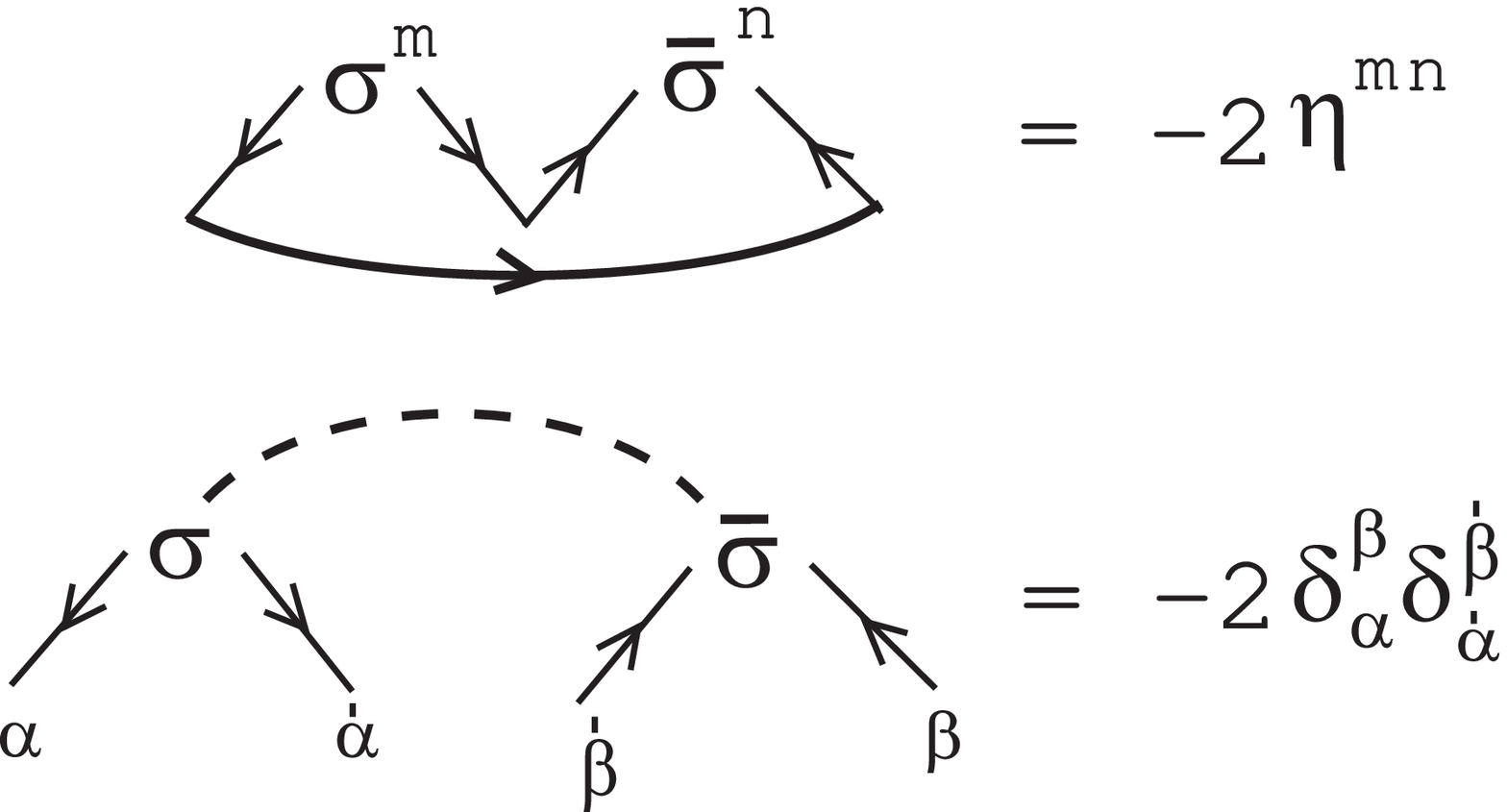}
            }
\label{LF13rev}
   \caption{ 
Completeness relations:\ 
1) $\del^\al_\be(\si^m)_{\al\aldot}(\sibar^n)^{\aldot\be}
=-2\eta^{mn}$,\ 
2) $(\si^m)_{\al\aldot}(\sibar_m)^{\bedot\be}
=-2\del^\be_\al\del^\bedot_\aldot$.\ 
The contraction using $\del^\al_\be$ is the matrix trace.
The relation 1) is also obtained from Fig.8.
   }
\end{figure}
The contraction, expressed by arrowed curve in Fig.10.
is the matrix trace.

Fierz identity is graphically shown as follows. 
\begin{eqnarray}
\graph{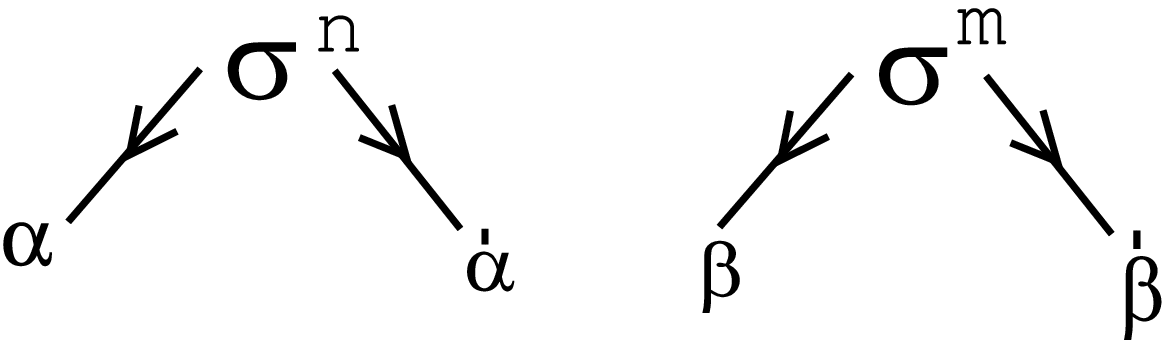}=\nn
\fourth\left\{
-\graph{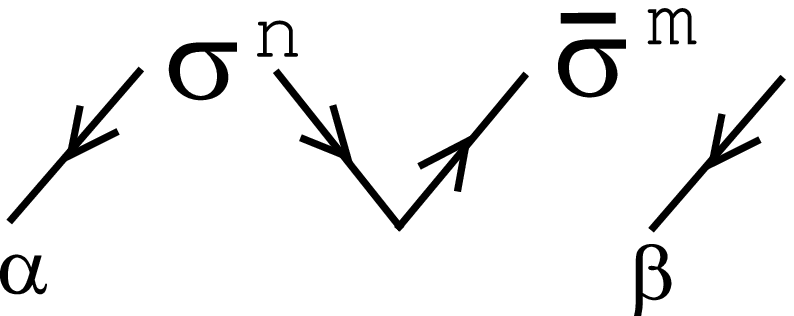}\ep_{\aldot\bedot}+\graph{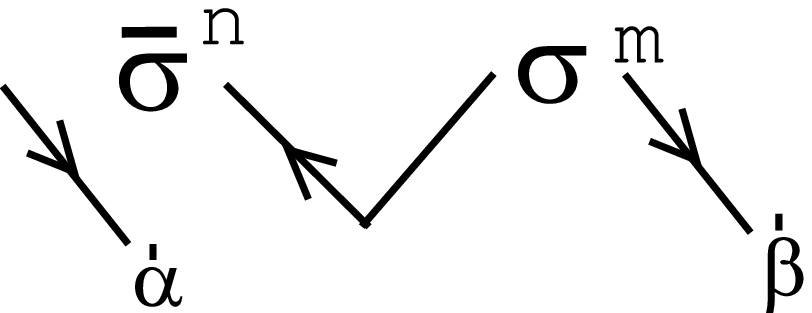}\ep_\ab
-m\change n\right\}\nn
-\half\eta^{nm}\ep_\ab \ep_{\aldot\bedot}
-\frac{1}{8}\left\{\graph{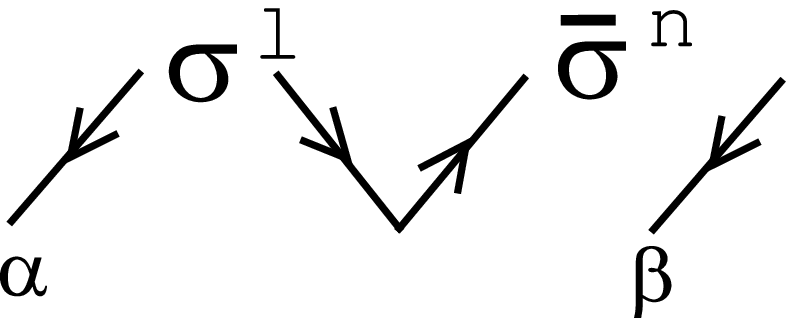}-l\change n\right\}
\left\{\graph{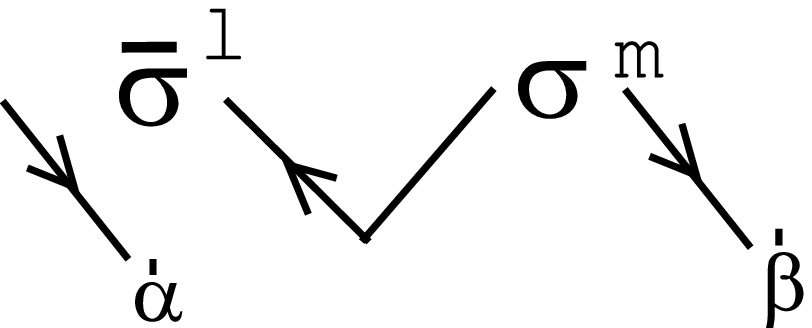}-l\change m\right\},
\label{def3}
\end{eqnarray}

These relations are important, in the program, to reduce
the product of sigma matrices. Especially at the place (E) of SigmaContraction, 
they are exploited. As an example, the closed chiral-loop graph in Fig.5 
reduces as follows.
\begin{eqnarray}
\graph{ssbssbBR}=
2(\eta^{lm}\eta^{ns}-\eta^{ln}\eta^{ms}+\eta^{ls}\eta^{mn}-i\ep^{lmns})
\label{ssbssbBR2}
\end{eqnarray}
Its complex conjugate one is given by
\begin{eqnarray}
\graph{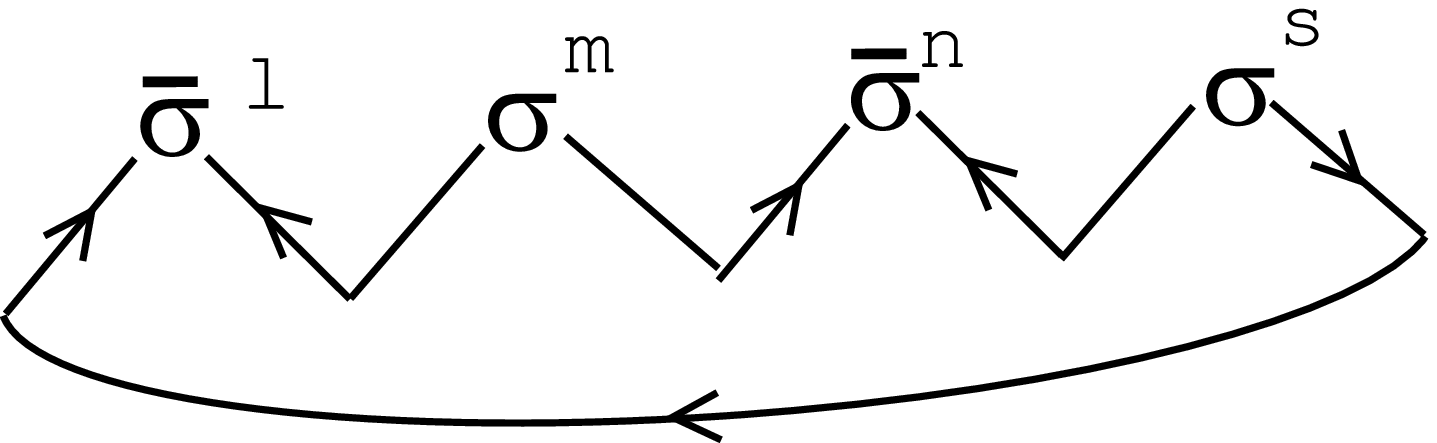}=
2(\eta^{lm}\eta^{ns}-\eta^{ln}\eta^{ms}+\eta^{ls}\eta^{mn}+i\ep^{lmns})
\label{sbssbsR}
\end{eqnarray}
The above 2 terms appear in the calculation of $W_\al W^\al$ and $\Wbar_\aldot \Wbar^\aldot$
respectively. ($W_\al$ and $\Wbar_\aldot$ are superfields of field strength. )

\section{Superfield and Structure of Input Data}\label{sec:sfield}
The transformation between
the superfield expression and the fields-components expression is an
important subject of SUSY theories. 
For the
purpose, we do the calculation of $\Phi^\dag\Phi$, in the App.B. In this
case, the input data is taken from the content of the superfield. The first
superfield $\Phi^\dag$(we assign the superfield-number sf=0) 
and the second one $\Phi$(sf=1) have 6 terms
(we assign the term-number t=0,1,$\cdots$,5) 
within each superfield. The input data, written in
App.B, can be read from the following expression.\nl

\begin{eqnarray}
\Phi^\dag=-i\graph{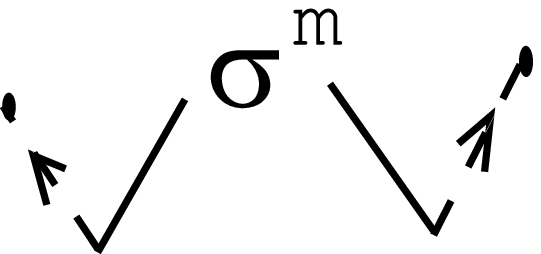}\pl_mA^*+\graph{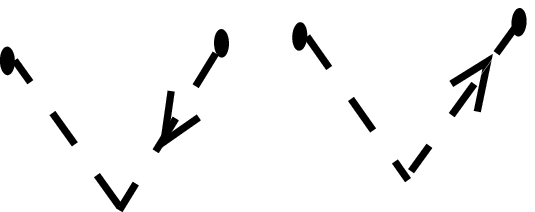}\fourth\pl^2A^*+2\graph{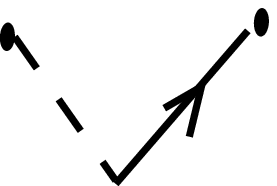}\nn
+i\graph{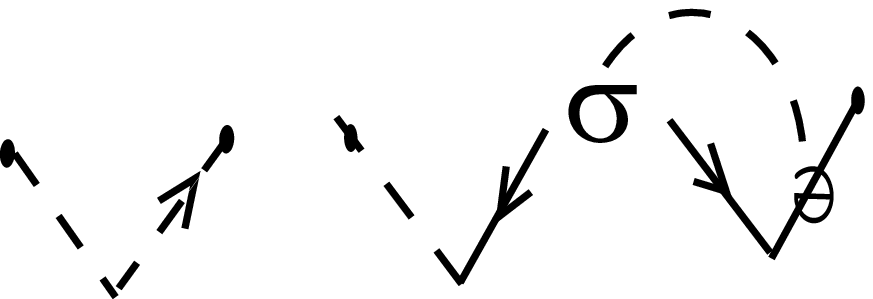}+\graph{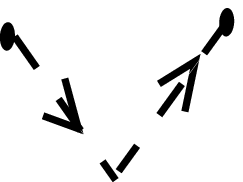}F^*+A^*
\label{PHIB}
\end{eqnarray}

This data of $\Phi^\dag$ is stored as

\shortstack[l]{
weight[sf=0,t=0]=0+i(-1)\\
type[sf=0,t=0,c=0]= t\\
th[sf=0,t=0,c=0,0,0]=1\\
type[sf=0,t=0,c=1]= s\\
si[sf=0,t=0,c=1,0,1]=1\\
si[sf=0,t=0,c=1,1,1]=2\\
siv[sf=0,t=0,c=1]=51\\
type[sf=0,t=0,c=2]= t\\
th[sf=0,t=0,c=2,1,0]=2\\
type[sf=0,t=0,c=3]= B\\
B[sf=0,t=0,c=3,1]=51
             }
\q
\shortstack[l]{
weight[sf=0,t=1]=1+i(0)\\
type[sf=0,t=1,c=0]= t\\
th[sf=0,t=1,c=0,0,0]=1\\
type[sf=0,t=1,c=1]= t\\
th[sf=0,t=1,c=1,0,1]=1\\
type[sf=0,t=1,c=2]= t\\
th[sf=0,t=1,c=2,1,1]=2\\
type[sf=0,t=1,c=3]= t\\
th[sf=0,t=1,c=3,1,0]=2\\
type[sf=0,t=1,c=4]= C\\
C[sf=0,t=1,c=4,1]=1
               }
\q$\cdot\cdot\cdot$\nl
\nl
Here we notice two additional coloumns, specified by sf and t, appear.
They specify the order of the supersields and the order of the terms
within each superfield respectively.

The superfield $\Phi$ is graphically shown as
\begin{eqnarray}
\Phi=i\graph{PHI1r}\pl_mA+\graph{PHI2r}\fourth\pl^2A+2\graph{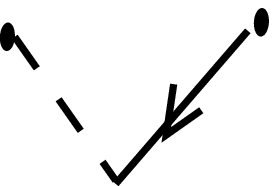}\nn
-i\graph{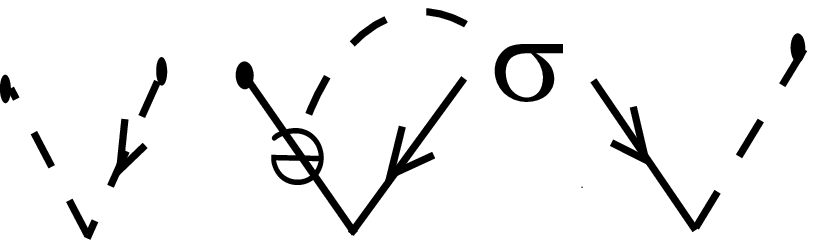}+\graph{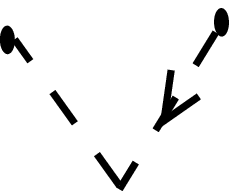}F+A
\label{PHI}
\end{eqnarray}

This data of $\Phi$ is stored as\nl
\shortstack[l]{
weight[sf=1,t=0]=0+i(1)\\
type[sf=1,t=0,c=0]= t\\
th[sf=1,t=0,c=0,0,0]=5\\
type[sf=1,t=0,c=1]= s\\
si[sf=1,t=0,c=1,0,1]=5\\
si[sf=1,t=0,c=1,1,1]=6\\
siv[sf=1,t=0,c=1]=52\\
type[sf=1,t=0,c=2]= t\\
th[sf=1,t=0,c=2,1,0]=6\\
type[sf=1,t=0,c=3]= B\\
B[sf=1,t=0,c=3,0]=52
               }
\q
\shortstack[l]{
weight[sf=1,t=1]=1+i(0)\\
type[sf=1,t=1,c=0]= t\\
th[sf=1,t=1,c=0,0,0]=5\\
type[sf=1,t=1,c=1]= t\\
th[sf=1,t=1,c=1,0,1]=5\\
type[sf=1,t=1,c=2]= t\\
th[sf=1,t=1,c=2,1,1]=6\\
type[sf=1,t=1,c=3]= t\\
th[sf=1,t=1,c=3,1,0]=6\\
type[sf=1,t=1,c=4]= C\\
C[sf=1,t=1,c=4,0]=1
               }
\q$\cdot\cdot\cdot$\nl

The calculation of $\Phi\dag\Phi$ leads to the Wess-Zumino Lagrangian. (See App.B)

We also do the calculation of $W_\al W^\al$ in App.B where $W_\al$ is the
field strength superfield. They are expressed as follows.\nl
\begin{eqnarray}
W_\al=
-i\graph{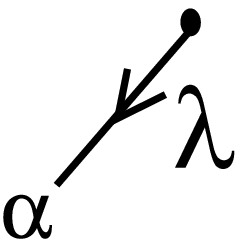}+\graph{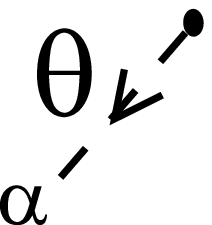}D-i\graph{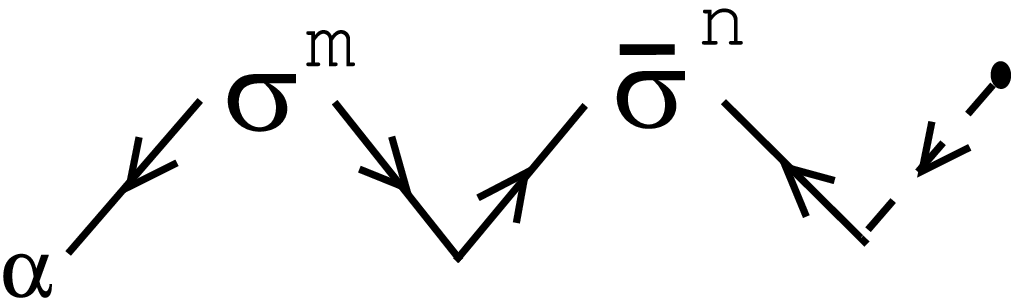}\half v_{mn}+\graph{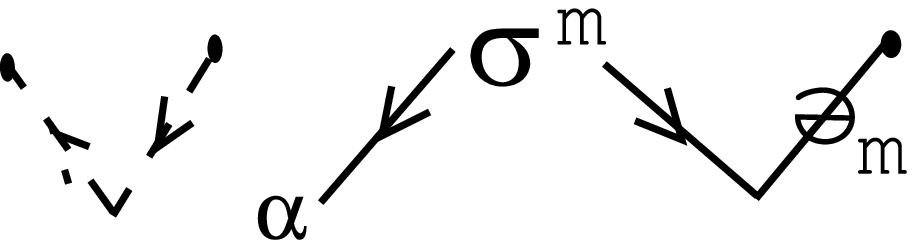}
\label{WalD}
\end{eqnarray}
This data of $W_\al$ is stored as

\shortstack[l]{
weight[sf=0,t=0]=0+i(-1)\\
type[sf=0,t=0,c=0]= l\\
la[sf=0,t=0,c=0,0,1]=1
               }
\q
\shortstack[l]{
weight[sf=0,t=1]=1+i(0)\\
type[sf=0,t=1,c=0]= t\\
th[sf=0,t=1,c=0,0,1]=1\\
type[sf=0,t=1,c=1]= D\\
D[sf=0,t=1,c=1]=1
               }
\q$\cdot\cdot\cdot$\nl

$W^\al$ is expressed as
\begin{eqnarray}
W^\al=
-i\graph{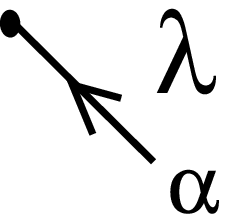}+\graph{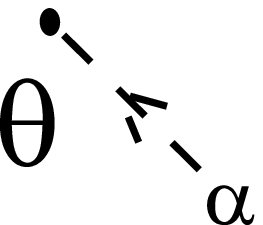}D-i\graph{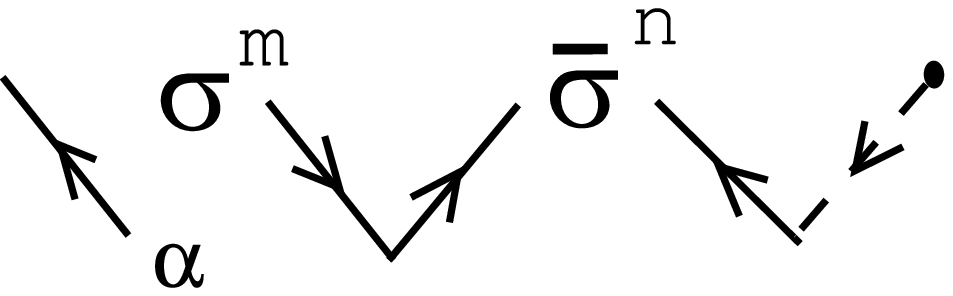}\half v_{mn}+\graph{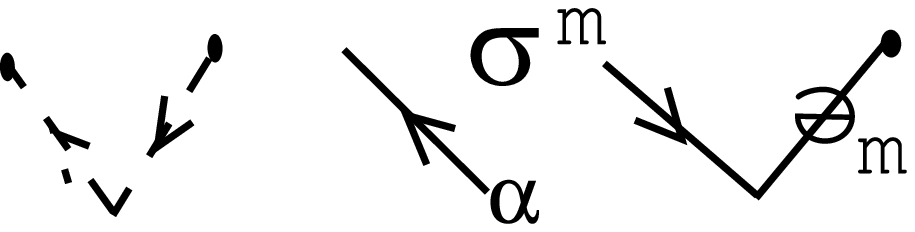}
\label{WalU}
\end{eqnarray}
This data of $W^\al$ is stored as\nl

\shortstack[l]{
weight[sf=1,t=0]=0+i(-1)\\
type[sf=1,t=0,c=0]= l\\
la[sf=1,t=0,c=0,0,0]=1
              }
\q
\shortstack[l]{
weight[sf=1,t=1]=1+i(0)\\
type[sf=1,t=1,c=0]= t\\
th[sf=1,t=1,c=0,0,0]=1\\
type[sf=1,t=1,c=1]= D\\
D[sf=1,t=1,c=1]=1
               }
\q$\cdot\cdot\cdot$\nl
The calculation of $W_\al W^\al$ leads to the Super Electromagnetism Lagrangian.(See App.B)

\section{Conclusion}\label{sec:conc}
In the history of the quantum field theory, 
new techniques have produced physically important results. 
The regularization techniques are such examples. 
The dimensional regularization by 'tHooft and Veltman\cite{TV72} produced
important results on the renormalization group property of Yang-Mills theory
and many scattering amplitude calculations. 
The lattice regularization in the gauge theory 
revealed non-perturbative features of hadron physics. 
In this case, the computor technique of numerical calculation
is essential. 
As for the computer algebraic one, we recall the calculation of 2-loop on-shell
counterterms of pure Einstein gravity\cite{GS85,Ven92}. 
A new technique is equally important as a new idea.
  
The SUSY theory is beautifully constructed respecting the symmetry
between bosons and fermions, but the  attractiveness
is practically much reduced by its complicated structure: many fields, 
chiral properties, Grassmannian algebra, etc.  
The present approach intends to improve
the situation by a computer program which makes use of the graphical technique. 
(This approach is taken in Ref.\cite{SI98IJMPC} for the calculation  of product of SO(N) tensors.
It was applied to various anomaly calculations. )

The present program should be much more improved. Here we cite the prospective
final goal.
\begin{enumerate}
\item
It can do the transformation between the superfield expression and
the component expression.
\item
It can do the SUSY trnasformation  
of various quantities. In particular it can confirm the SUSY-invariance 
of the Lagrangian in the graphical way and give the final total divergence.
\item
It can do algebraic SUSY calculation involving $D_\al, \Dbar^\aldot, Q_\al$ and $\Qbar^\aldot$.
\end{enumerate}
The item 1 above has been demostrated in the present paper.

It is impossible to deal with all SUSY calculations. This is simply because
which fields appear and which dimensional quantities are calculated
depend on each problem. If we obatin a list of (graph) indices which classify
all phsical quantities (operators) appearing in the output, then the present
program works (by adding new lines for the new problem). To deal with
such a case, we add the appendix B where the program flow is explained. It intends
to help the reader to read the original source code. 

\section{Appendix A: Skech of Programming Flow}\label{sec:appA}
\footnote{
For simplicity, we omit the vector multiplet components.
}
\subsection{Input of Data}
\setlength{\unitlength}{1mm}
\begin{picture}(150,200)(0,0)
\put(0,130){\framebox(150,70)[tl]{     
\shortstack[l]{
[Initialization of Basic Components]\\
sf=0,1,$\cdots$maxsfieldno-1;\ t=0,1,$\cdots$maxtermno-1;\ 
c=0,1,$\cdots$maxcompno-1;\ i,j,k=0,1\\
type[sf][t][c]='x'\\
th[sf][t][c][i][j]=99\\
psi[sf][t][c][i][j]=99\\
dps[sf][t][c][i][j]=99\\
dpsv[sf][t][c]=99\\
si[sf][t][c][i][j]=99\\
siv[sf][t][c]=99\\
A[sf][t][c][k]=99\\
B[sf][t][c][k]=99\\
C[sf][t][c][k]=99\\
F[sf][t][c][k]=99
             }
}}                                        
\put(-5,80){\makebox(150,50)[tl]{     
\shortstack[l]{
\quad sfieldno=2\\
\quad for (sf=0;sf<sfieldno;sf++)\\
\{  /* <sf-running */  \\
\quad printf("SUPER FIELD  NO" ,sf)\\
\quad fscanf (\& termno)$\longrightarrow$ printf,\ noterm[sf]\\
\\
\quad for (t=0;t<termno;t++)\\
\{  /* <t-running */  \\
\quad printf("term=",t)
             }
}}                                     

\put(0,60){\framebox(150,15)[tl]{     
\shortstack[l]{
[Initialization of Graph Indices]\\
dif=0,\ nth=thn=thbarn=0,\ npsi=psin=psibarn=0,\ 
nsi=sin=sibarn=0\\
nA=An=Abarn=0,\ nB=Bn=Bbarn=0,\ 
nC=Cn=Cbarn=0,\ nF=Fn=Fbarn=0
             }
}}                                        
\nl
\put(-5,0){\makebox(150,50)[tl]{     
\shortstack[l]{
\quad fscanf ({\&}weight[sf][t][2])$\longrightarrow$ printf\\
\quad fscanf ({\&}compno)$\longrightarrow$ printf,\ nocomp[sf][t]\\
\q\\
for (c=0;c<compno;c++)\\
\{  /* <c-running */  \\
\quad printf(c)\\
\quad fscanf ({\&}type[sf][t][c])$\longrightarrow$ printf
             }
}}                                     
\put(30,20){\vector(0,-1){10}}

\put(20,0){\framebox(30,10)[tl]{     
\shortstack[l]{SWITCH
             }
}}                                        
\end{picture}

\newpage

[Install of SuperField Data]\nl
\begin{picture}(150,200)(0,0)
\put(0,150){\makebox(150,50)[tl]{     
\shortstack[l]{
case 't' \qq       /* Spinor Coordinate $\sh,\thbar$ */ \\
\qqq nth$\ra$ nth+1\q\q printf(nth)\\
\qqq fscanf({\&}lr,{\&}ud, {\&}spsuf)\\
\qqq switch(lr)$\ra$ 
$\left\{
\begin{array}{l}
\mbox{lr=0\q thn} \ra \mbox{thn+1\q printf(thn)}\\
\mbox{lr=1\q thbarn} \ra \mbox{thbarn+1\q printf(thbarn)}
\end{array}
\right.$ 
\\
\qqq printf (spsuf)\\
\qqq th[sf][t][c][lr][ud]=spsuf
             }
}}                                     
\put(0,100){\makebox(150,50)[tl]{     
\shortstack[l]{
case 'p' \qq       /* Spinor $\psi,\psibar$ */ \\
\qqq npsi$\ra$ npsi+1\q\q printf(npsi)\\
\qqq fscanf({\&}lr,{\&}ud, {\&}spsuf)\\
\qqq switch(lr)$\ra$ 
$\left\{
\begin{array}{l}
\mbox{lr=0\q psin} \ra \mbox{psin+1\q printf(psin)}\\
\mbox{lr=1\q psibarn} \ra \mbox{psibarn+1\q printf(psibarn)}
\end{array}
\right.$ 
\\
\qqq printf (spsuf)\\
\qqq psi[sf][t][c][lr][ud]=spsuf
             }
}}                                     
\put(0,50){\makebox(150,50)[tl]{     
\shortstack[l]{
case 'q' \qq       /* der Spinor $\pl\psi,\pl\psibar$ */ \\
\qqq npsi$\ra$ npsi+1,\q dif$\ra$dif+1\q\q printf(npsi)\\
\qqq fscanf({\&}lr,{\&}ud, {\&}spsuf)\\
\qqq switch(lr)$\ra$
$\left\{
\begin{array}{l}
\mbox{lr=0\q psin} \ra \mbox{psin+1\q printf(psin)}\\
\mbox{lr=1\q psibarn} \ra \mbox{psibarn+1\q printf(psibarn)}
\end{array}
\right.$ 
\\
\qqq printf (spsuf)\\
\qqq dps[sf][t][c][lr][ud]=spsuf\\
\qqq fscan({\&}vecsuf)\\
\qqq dpsv[sf][t][c]=vecsuf
             }
}}                                     
\end{picture}

\newpage

\begin{picture}(150,200)(0,0)
\put(0,130){\makebox(150,70)[tl]{     
\shortstack[l]{
case 's' \qq       /* sigma $\si,\sibar$ */ \\
\qqq nsi$\ra$ nsi+1\q\q printf(nsi)\\
\qqq fscanf({\&}lr1,{\&}ud1, {\&}spsuf1)\\
\qqq fscanf({\&}lr2,{\&}ud2, {\&}spsuf2)\\
\qqq\q
$\left\{
\begin{array}{l}
\mbox{lr1=0,lr2=1 (case'901')\q sin} \ra \mbox{sin+1, printf}\\
\mbox{lr1=1,lr2=0 (case'910')\q sibarn} \ra \mbox{sibarn+1, printf}
\end{array}
\right.$ 
\\
\qqq si[sf][t][c][lr1][ud1]=spsuf1\\
\qqq si[sf][t][c][lr2][ud2]=spsuf2\\
\qqq fscanf({\&}vecsuf)\\
\qqq siv[sf][t][c]=vecsuf
             }
}}                                     
\put(0,80){\makebox(150,50)[tl]{     
\shortstack[l]{
case 'B' \qq       /* der Scalar $\pl A,\pl\Abar$ */ \\
\qqq nB$\ra$ nB+1, dif$\ra$dif+1\q\q printf(nB)\\
\qqq fscanf({\&}lr, {\&}vecsuf)\\
\qqq\q
$\left\{
\begin{array}{l}
\mbox{lr=0 ($\pl A$)\q Bn} \ra \mbox{Bn+1, printf(Bn)}\\
\mbox{lr=1 ($\pl A^{*}$)\q Bbarn} \ra \mbox{Bban+1, printf(Bbarn)}
\end{array}
\right.$ 
\\
\qqq B[sf][t][c][lr]=vecsuf,\q printf
             }
}}                                     
\put(0,30){\makebox(150,50)[tl]{     
\shortstack[l]{
case 'A' \qq       /* Scalar $A,\Abar$ */ \\
\qqq nA$\ra$ nA+1,\q printf(nA)\\
\qqq fscanf({\&}lr, {\&}existno)\\
\qqq\q
$\left\{
\begin{array}{l}
\mbox{lr=0 ($A$)\q An} \ra \mbox{An+1, printf(An)}\\
\mbox{lr=1 ($A^{*}$)\q Abarn} \ra \mbox{Abarn+1, printf(Abarn)}
\end{array}
\right.$ 
\\
\qqq A[sf][t][c][lr]=existno,\q printf
             }
}}                                     

\put(0,0){\makebox(150,30)[tl]{     
\shortstack[l]{
case 'C' \qq       /* der de Scalar $\pl\pl A,\pl\pl\Abar$ */ \\
\qqq nC$\ra$ nC+1,\q dif$\ra$ dif+2,\q printf(nC)\\
\qqq fscanf({\&}lr, {\&}existno)\\
\qqq\q
$\left\{
\begin{array}{l}
\mbox{lr=0 ($\pl\pl A$)\q Cn} \ra \mbox{Cn+1, printf(Cn)}\\
\mbox{lr=1 ($\pl\pl A^{*}$)\q Cbarn} \ra \mbox{Cbarn+1, printf(Cbarn)}
\end{array}
\right.$ 
\\
\qqq C[sf][t][c][lr]=existno,\q printf
             }
}}                                     

\end{picture}
\newpage

\} /* c-running> */\nl
\} /* t-running> */\nl
\} /* sf-running> */\nl

/* Expanding the product of superfields */\nl
\framebox(150,25)[tl]{     
\shortstack[l]{
T[0]=0,1,$\cdots$,noterm[0]-1:\ Term No of Superfield 0\\
T[1]=0,1,$\cdots$,noterm[1]-1:\ Term No of Superfield 1\\
$\cdot$\\
$\cdot$\\
T[sfieldno-1]=0,1,$\cdots$,noterm[sfieldno-1]-1:\ Term No of Superfield sfieldno-1
                }
                      }
 For every set (T[0],T[1],$\cdots$,T[sfieldno-1]), do the following box.\nl
\vspace{3mm}
printf (T[0],T[1],$\cdots$,T[sfieldno-1])\nl
\framebox(150,60)[tl]{     
\shortstack[l]{
\\
termscombine(); \\
\\
thn=THnum1[0]; thbarn=THnum1[1];\\
if(thn >= 3) \{ printf("NO of theta: MORE THAN 2"); goto lab41; \} \\
if(thbarn >= 3) \{ printf("NO of theta-bar: MORE THAN 2"); goto lab41; \} \\
\\
SORTOUTthBth();\\
\\
thBthBthth();\\
\\
SigmaContraction();\\
\\
VecSufChange();\\
\\
lab41:; 
             }
}\nl
\nl
/* thn:\ No of $\sh$ within a term;\ thbarn:\ No of $\thbar$ within a term;\   */ \nl                                      
\subsection{termscombine}
printf("*** TERMSCOMBINE ***");\nl
{\framebox(80,70)[tl]{     
\shortstack[l]{
[Initialization]\\
c=0,1,$\cdots$,maxcompno2-1\\
type1[c]='x'\\
th1[c][ ][ ]=99\\
si1[c][ ][ ]=99\\
dpsv1[c]=99\\
psi1[c][ ][ ]=99\\
dps1[c][ ][ ]=99\\
B1[c][ ]=99\\
A1[c][ ]=99\\
F1[c][ ]=99\\
C1[c][ ]=99
             }
}}
\nl\nl
(T(0)term of sf=0 SuperField)$\times$\nl                                        
(T(1)term of sf=1 SuperField)$\times$\nl                                        
$\cdot$\nl
$\cdot$\nl
(T(sfieldno-1)term of sf=sfieldno-1 SuperField)\nl                                        
\nl
\q\q $\longrightarrow$ One Term Combined\nl
\nl
Indices of a combined term\nl
"Gathered Term Check"\nl
No of Components, weight1[0], weight1[1], DIFnum1, THnum1[ ], SIGnum[ ],
PSInum1[ ], Bnum[ ], Anum1[ ], Fnum1[ ], Cnum1[ ]\nl

\subsection{SORTOUTthBth}
{\it Moving} th($\theta$), thbar($\thbar$), si($\si$) and sibar($\sibar$)
{\it to the biggining part of a term}. In the process, we obtain
the sign change due to the Grassman algebra.\nl
\nl
{\framebox(80,5)[tl]{     
\shortstack[l]{
PlusMinus\ :\ Sign change
             }
}}
\nl
Here we use the function: idno(TP). 

idno(TP);\ a function which produces FieldID number correponding
to a type TP. 

\subsection{thBthBthth}
First we change the data form of th[\ ][\ ][\ ]. From :\nl
\framebox(50,10)[tl]{     
\shortstack[l]{
th[c][i][j]=$\al$,\ i,j=0 or 1\\
c=0,1,$\cdots$,nth-1 }
                     }
\nl
to an another form:\nl
\framebox(50,7)[tl]{     
\shortstack[l]{
I[c]=i,\ J[c]=j,\ Suf[c]=$\al$     
              }
                     } 

In this part, the program classifies terms by the power of $\sh$ and $\thbar$.\nl 
(1) indipendent of $\sh$ and $\thbar$ (nth=0)\nl
(2) $\sh^\al, \sh_\al, \thbar_\aldot, \thbar^\aldot$ (nth=1)\nl
(3) $\sh^\al\thbar^\bedot, \sh^\al\thbar_\bedot, \cdots$ (nth=2)\nl
(4) $\sh^\al\sh^\be, \sh^\al\sh_\be, \cdots; 
      \thbar^\aldot\thbar^\bedot, \thbar^\aldot\thbar_\bedot, \cdots; \sh^2; \thbar^2$ (nth=2)\nl
(5) $\sh^2(\thbar_\aldot, \thbar^\aldot), \thbar^2(\sh_\al, \sh^\al)$ (nth=3)\nl
(6) $\sh^\al\sh^\be\thbar_\gadot,\cdots; \thbar^\aldot\thbar^\bedot\sh_\ga,\cdots$ (nth=3)\nl
(7) $\sh^\al\thbar^\bedot\thbar^\gadot\sh^\del,\cdots;
\thbar^\aldot\sh^\be\sh^\ga\thbar^\deldot,\cdots; \cdots$ (nth=4)\nl
(8) $\sh^\al\thbar^\bedot\sh^\ga\thbar^\gadot, \cdots; \thbar^\aldot\sh^\be\thbar^\gadot\sh^\del, 
\cdots; \cdots$ (nth=4)\nl
(9) $\sh^\al\sh^\be\thbar^\gadot\thbar^\deldot,\cdots; \thbar^\aldot\thbar^\bedot\sh^\ga\sh^\del,
\cdots;\cdots$ (nth=4)\nl
\nl
The formula presented in Sec.\ref{sec:scoordinate} is exploited here. 
As the output, we obtain\nl
{\framebox(95,10)[tl]{     
\shortstack[l]{
\q [OutPut]\\
weight,\ PlusMinus,\ Sign,\ Nthth,\ NthBthB,\ Nhalf
             }
}}
\nl
This OutPut means the following factor
$$
\mbox{(weight)}*(-1)^{\mbox{PlusMinus}+\mbox{Sign}}*
(\frac{1}{2})^{\mbox{Nhalf}}*(\sh^2)^{\mbox{Nthth}}
(\thbar^2)^{\mbox{NthBthB}}
$$
appears in the process of this subsection. 'weight' is the starting one:\nl 
weight=weight[0]+i~weight[1]. 
Here we use SufChange().\nl
\nl
SufChange():\ Chiral suffix contraction using $\del_\al^\be$, $\ep^{\al\be}$, 
$\ep_{\al\be}$. It is done for component c with c$\geq$nth or c$\geq$nth+nsi.
\nl
\nl

\subsection{SigmaContraction}
This routine is composed of 5 parts, (A),(B),(C),(D) and (E). \nl

(A) Change of the data form \nl
First we change the data form:\ (N=nsi-1)\nl
\shortstack[l]{
$\si^{m_0}_{\al_0\aldot_0}$\\
{\framebox(37,10)[tl]{     
\shortstack[l]{
si[nth][2][2]\\
siv[nth]      }
}}                                        
             }  
\q
\shortstack[l]{
$\si^{m_1\al_1\aldot_1}$\\
{\framebox(37,10)[tl]{     
\shortstack[l]{
si[nth+1][2][2]\\
siv[nth+1]      }
}}                                        
             }  
\q$\cdots$\q
\shortstack[l]{
$\si^{m_N\ \aldot_N}_{\ \ \al_N}$\\
{\framebox(37,10)[tl]{     
\shortstack[l]{
si[nth+nsi-1][2][2]\\
siv[nth+nsi-1]      }
}}                                        
             }  
\nl
to an another form:\nl
\shortstack[l]{
c=0\\
{\framebox(37,25)[tl]{     
\shortstack[l]{
Csuff[0][0]=$\al_0$\\
Csuff[0][1]=$\aldot_0$\\
Vsuff[0]=m$_0$\\
UD[0][0]=1(down)\\
UD[0][1]=1(down)     
              }
}}                                        
             }  
\q
\shortstack[l]{
c=1\\
{\framebox(37,25)[tl]{     
\shortstack[l]{
Csuff[1][0]=$\al_1$\\
Csuff[1][1]=$\aldot_1$\\
Vsuff[1]=m$_1$\\
UD[1][0]=0(up)\\
UD[1][1]=0(up)     
              }
}}                                        
             }  
\q
$\cdots$
\q
\shortstack[l]{
c=N\\
{\framebox(37,25)[tl]{     
\shortstack[l]{
Csuff[N][0]=$\al_N$\\
Csuff[N][1]=$\aldot_N$\\
Vsuff[N]=m$_N$\\
UD[N][0]=1(down)\\
UD[N][1]=0(up)     
              }
}}                                        
             }  \nl

  (B) Chiral-suffix-pair search\nl
Search for same suffixes in \{Csuff[c][0]|c=0,1,$\cdots$,N\}(odd number suffixes) and in
\{Csuff[c][1]|c=0,1,$\cdots$,N\}(even number suffixes). When Csuff[c1][0]=Csuff[c2][0], then
we call (c1,c2) chiral-suffix-pair (or, simply, Cpair). For the odd number case, we list them as\nl
\nl
{\framebox(95,55)[tl]{     
\shortstack[l]{
CpairO[cpairnoO=0][0]=c1\\
CpairO[cpairnoO=0][1]=c2   (c1<c2)\\
CpairO[cpairnoO=0][2]=$\al_0$(common chiral suffix)\\
CpairO[cpairnoO=1][0]=c3\\
CpairO[cpairnoO=1][1]=c4   (c3<c4)\\
CpairO[cpairnoO=1][2]=$\al_1$\\
\q\q$\cdot$\\
\q\q$\cdot$\\
CpairO[NcpairO-1][0]=c*\\
CpairO[NcpairO-1][1]=c+   (c*<c+)\\
CpairO[NcpairO-1][2]=$\al_{NcpairO-1}$
              }
}}                                        
\nl
  For the even number case, \nl
\nl
{\framebox(100,55)[tl]{     
\shortstack[l]{
CpairE[cpairnoE=0][0]=c10\\
CpairE[cpairnoE=0][1]=c11   (c10<c11)\\
CpairE[cpairnoE=0][2]=$\aldot_0$(common anti-chiral suffix)\\
CpairE[cpairnoE=1][0]=c12\\
CpairE[cpairnoE=1][1]=c13   (c12<c13)\\
CpairE[cpairnoE=1][2]=$\aldot_1$\\
\q\q$\cdot$\\
\q\q$\cdot$\\
CpairE[NcpairE-1][0]=c**\\
CpairE[NcpairE-1][1]=c++    (c**<c++)\\
CpairE[NcpairE-1][2]=$\aldot_{NcpairE-1}$
              }
}}                                        
\nl

(C) Grouping $\si$'s\nl
A group of $\si$'s is defined to be a set of $\si$'s which are
connected by chiral-suffix or anti-chiral-suffix contractions.
The number of groups within a term is assigned to be {\bf GrNum}. 
This an important Graph index. 
The number of $\si$'s which each group has is stored in {\bf SigmaN}[GR]\ 
(GR=0,1,$\cdots$,GrNum-1). Each group consits of some $\si$'s specified their component numbers. 
They are stored in {\bf Group}[GR][nseries]\ (nseries=0,1,$\cdot$,SigmaN[GR]-1). In this process
of grouping, the program traces the CpairO and CpairE defined in (B). Each group is 
determined by a series of cpairnoO's and cpainoE's. They are stored in {\bf CPseries}[GR][\ ]. \nl

(D)   Vector-suffix-pair search\nl
Search the same vector-suffix in \{Vsuff[c]|c=0,1,$\cdots$,N\}. \nl
{\framebox(100,25)[tl]{     
\shortstack[l]{
Vpair[0][0]=c1,\ Vpair[0][1]=c2\\
Vpair[1][0]=c3,\ Vpair[1][1]=c4\\
\q\q$\cdot$\q\q\q$\cdot$\\
\q\q$\cdot$\q\q\q$\cdot$\\
Vpair[vpairno-1][0]=c*,\ Vpair[vpairno-1][1]=c+
              }
}}                                        
\nl

(E) Classifying $\si$'s part by indices (nsi, vpairno, NcpairO, NcpairE, closed chiral-loop)

Using graph indices defined in the Sec.\ref{sec:gind},
we can classify $\si$'s part. See the TABLE 2,3 and 4 in the text. 
Each class (we call here the finally classified place "class") has the
number SigGraN which is the set of the graph indices. In each class,
we reduce the expression using 
the graphical relations explained in Sec.\ref{sec:sigma}. 
Here the following important data are obtained and transfered to VecSufChange(). 

{\bf BranchN}:\ Generally each class decomposes to some branches after the reduction using
the graphical relations. We specify the number of branches.\nl 
{\bf NoChangeN[br]}:\ For each branch(br), NoChangeN is assigned, 0 for the case that $\si$'s part
does not change, 1 for the case that $\si$'s part changes.\nl 
{\bf RedSigN[br]}:\ For each branch, the decrease-number of $\si$'s after the reduction
is stored. This number is important for adjusting the output form.\nl
{\bf cstart}:\ Normally this is fixed as cstart=nth+nsi\nl
{\bf LorenContN[br]}:\ For each branch, specify the number of Lorenz contraction, in other words,
the number of $\eta^{mn}$.\nl 
{\bf SUFF[br][10][2]}:\ Specify the vector suffixes appearing in $\eta$'s.\nl  
{\bf MultiFac[br][2]}:\ For each branch, a multiplicative factor is specified.\nl 
{\bf si1BR[br][c][2][2], siv1BR[br][c]}:\ For each branch, specify the $\si$'s part.\nl 
{\bf ep1[c][4]}:\ For each branch, specify the totally-antisymmetric tensor when
this term appears. \nl

\subsection{VecSufChange}

Vector suffix contraction is peformed using $\eta^{mn}\eta^{ls}\cdots$. 
It is done for component c (c$\geq$cstart=nth+nsi).

\subsubsection{FinalOutPut}

\section{Appendix B: Input and Output Examples}\label{sec:appB}
\subsection{Wess-Zumino Model}
We demonstrate the calculation of $\Phi^\dag \Phi$, where $\Phi$ is 
the chiral superfield, and obtain the component form. The next input data
can be read from the graphical eqauations (\ref{PHIB}) and (\ref{PHI}). 
\nl
\nl
{\bf INPUT DATA}\nl

2

6

0 -1\q
4\q
t\q
0 0 1\q
s\q
0 1 1\q
1 1 2\q
51\q
t\q
1 0 2\q
B\q
1 51\q

1 0\q
5\q
t\q
0 0 1\q
t\q
0 1 1\q
t\q
1 1 2\q
t\q
1 0 2\q
C\q
1 1\q

2 0\q
2\q
t\q
1 1 2\q
p\q
1 0 2\q

0 1\q
5\q
t\q
1 1 2\q
t\q
1 0 2\q
t\q
0 0 1\q
s\q
0 1 1\q
1 1 4\q
51\q
q\q
1 0 4\q
51\q

1 0\q
3\q
t\q
1 1 2\q
t\q
1 0 2\q
F\q
1 1\q

1 0\q
1\q
A\q
1 1\q

6

0 1\q
4\q
t\q
0 0 5\q
s\q
0 1 5\q
1 1 6\q
52\q
t\q
1 0 6\q
B\q
0 52\q

1 0\q
5\q
t\q
0 0 5\q
t\q
0 1 5\q
t\q
1 1 6\q
t\q
1 0 6\q
C\q
0 1\q

2 0\q
2\q
t\q
0 0 5\q
p\q
0 1 5\q

0 -1\q
5\q
t\q
0 0 5\q
t\q
0 1 5\q
q\q
0 0 7\q
52\q
s\q
0 1 7\q
1 1 6\q
52\q
t\q
1 0 6\q

1 0\q
3\q
t\q
0 0 5\q
t\q
0 1 5\q
F\q
0 1\q

1 0\q
1\q
A\q
0 1
\nl
\nl
{\bf OUT PUT}\nl

T[0]=0  T[1]=0\nl  

****** TERMSCOMBINE ****\nl
th*th*thbar*thbar-term\nl

****** SORTOUTthBth ****\nl
lab50 at thBthBthth, Final result\nl
weight= 1+i(0)\q
PlusMinus= 0\q
Sign=2\q
Nthth=1\q
NthBthB=1\q
Nhalf=2\nl

******  SigmaContraction  ****\nl
SigGraN=201199\q
FinalOutPut: MultiFac=-2 + i(0)\q
type2[c=6]= B\q
B2[c=6,1]=51\q
type2[c=7]= B\q
B2[c=7,0]=51\nl

T[0]=1  T[1]=5 \nl 

****** TERMSCOMBINE ****\nl
th*th*thbar*thbar-term\nl

****** SORTOUTthBth ****\nl
lab50 at thBthBthth, Final result\nl
weight= 1+i(0)\q
PlusMinus= 0\q
Sign=0\q
Nthth=1\q
NthBthB=1\q
Nhalf=0\nl

******  SigmaContraction  ****\nl
SigGraN=0\q
FinalOutPut: MultiFac=1 + i(0)\q
type2[c=4]= C\q
C2[c=4,1]=1\q
type2[c=5]= A\q
A2[c=5,0]=1\nl

T[0]=2  T[1]=3\nl  

****** TERMSCOMBINE ****\nl
th*th*thbar*thbar-term\nl

****** SORTOUTthBth ****\nl
lab50 at thBthBthth, Final result\nl 
weight= 0+i(-2)\q
PlusMinus= 4\q
Sign=0\q
Nthth=1\q
NthBthB=1\q
Nhalf=1\nl

******  SigmaContraction  ****\nl
SigGraN=1\q
FinalOutPut: MultiFac=1 + i(0)\q
type2[c=4]= s\q
si2[c=4,0,1]=7\q
si2[c=4,1,1]=2\q
siv2[c=4]=52\q
type2[c=5]= p\q
psi2[c=5,1,0]=2\q
type2[c=6]= q\q
dps2[c=6,0,0]=7\q
dpsv2[c=6]=52\nl

T[0]=3  T[1]=2\nl  

****** TERMSCOMBINE ****\nl
th*th*thbar*thbar-term\nl

****** SORTOUTthBth ****
lab50 at thBthBthth, Final result\nl 
weight= 0+i(2)\q
PlusMinus= 1\q
Sign=1\q
Nthth=1\q
NthBthB=1\q
Nhalf=1\nl

******  SigmaContraction  ****\nl
SigGraN=1\q
FinalOutPut: MultiFac=1 + i(0)\q
type2[c=4]= s\q
si2[c=4,0,1]=1\q
si2[c=4,1,1]=4\q
siv2[c=4]=51\q
type2[c=5]= q\q
dps2[c=5,1,0]=4\q
dpsv2[c=5]=51\q
type2[c=6]= p\q
psi2[c=6,0,0]=1\nl

T[0]=4  T[1]=4\nl  

****** TERMSCOMBINE ****\nl
th*th*thbar*thbar-term\nl

****** SORTOUTthBth ****\nl
lab50 at thBthBthth, Final result \nl
weight= 1+i(0)\q
PlusMinus= 0\q
Sign=0\q
Nthth=1\q
NthBthB=1\q
Nhalf=0\nl

******  SigmaContraction  ****\nl
SigGraN=0\q
FinalOutPut: MultiFac=1 + i(0)\q
type2[c=4]= F\q
F2[c=4,1]=1\q
type2[c=5]= F\q
F2[c=5,0]=1\nl

T[0]=5  T[1]=1\nl  

****** TERMSCOMBINE ****\nl
th*th*thbar*thbar-term\nl

****** SORTOUTthBth ****\nl
lab50 at thBthBthth, Final result\nl 
weight= 1+i(0)\q
PlusMinus= 0\q
Sign=0\q
Nthth=1\q
NthBthB=1\q
Nhalf=0\nl

******  SigmaContraction  ****\nl
SigGraN=0\q
FinalOutPut: MultiFac=1 + i(0)\q
type2[c=4]= A\q
A2[c=4,1]=1\q
type2[c=5]= C\q
C2[c=5,0]=1\nl

Gathering all terms, we obtain
\begin{eqnarray}
\Phi^\dag\Phi|_{\sh^2\thbar^2}==-\half \pl_mA^*\pl^mA\ {\bf (0,0)}
+\fourth\pl^2A^*\cdot A\ {\bf (1,5)}+\fourth A^*\pl^2A\ {\bf (5,1)}\nn
-i\times(-1)\graph{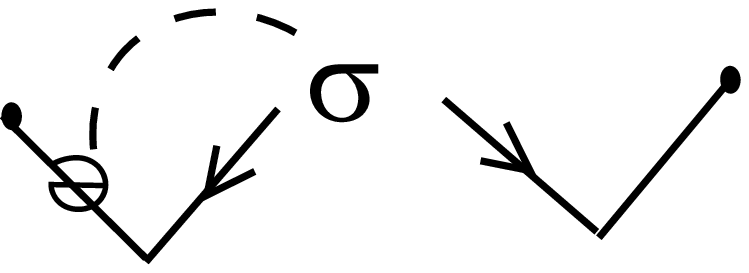}\ {\bf (2,3)}+i\times (-1)\graph{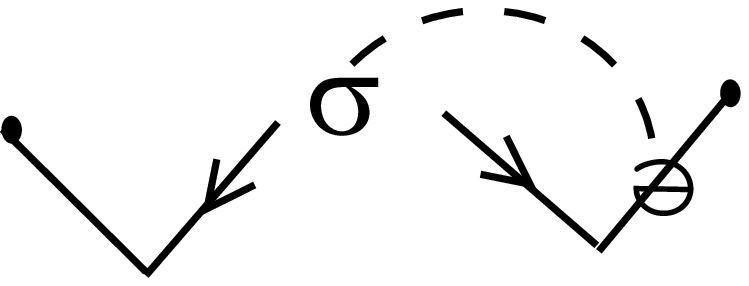}\ {\bf (3,2)}
+F^*F\ {\bf (4,4)}
\label{PHIBPHI}
\end{eqnarray}
This is the Wess-Zumino Lagrangian. We donot ignore the total divergence here.

\subsection{Super QED}
The calculation of $W_\al W^\al$ gives the kinetic terms of the photon and the photino
in Super QED. 
The following input data
can be read from the graphical eqauations (\ref{WalD}) and (\ref{WalU}). 
\nl
\nl
{\bf INPUT DATA}\nl

2

4

0 -1\q
1\q
l\q
0 1 1\q

1 0\q
2\q
t\q
0 1 1\q
D\q
1\q

0 -1\q
4\q
s\q
0 1 1\q
1 1 2\q
51\q
s\q
0 0 3\q
1 0 2\q
52\q
t\q
0 1 3\q
w\q
52\q
51\q

1 0\q
4\q
t\q
0 0 3\q
t\q
0 1 3\q
s\q
0 1 1\q
1 1 2\q
51\q
m\q
1 0 2\q
51\q

4

0 -1\q
1\q
l\q
0 0 1\q

1 0\q
2\q
t\q
0 0 1\q
D\q
1\q

0 -1\q
4\q
s\q
0 0 1\q
1 1 4\q
53\q
s\q
0 0 5\q
1 0 4\q
54\q
t\q
0 1 5\q
w\q
54\q
53\q

1 0\q
4\q
t\q
0 0 5\q
t\q
0 1 5\q
s\q
0 0 1\q
1 1 4\q
53\q
m\q
1 0 4\q
53
\nl
\nl
{\bf OUT PUT}\nl

T[0]=0  T[1]=3\nl  

****** TERMSCOMBINE ****\nl
th*th-term\nl

****** SORTOUTthBth ****\nl
lab50 at thBthBthth, Final result\nl 
weight= 0+i(-1)\q
PlusMinus= 2\q
Sign=0\q
Nthth=1\q
NthBthB=0\q
Nhalf=0\nl

******  SigmaContraction  ****\nl
SigGraN=1\q
FinalOutPut: MultiFac=1 + i(0)\q
type2[c=2]= s\q
si2[c=2,0,0]=1\q
si2[c=2,1,1]=4\q
siv2[c=2]=53\q
type2[c=3]= l\q
la2[c=3,0,1]=1\q
type2[c=4]= m\q
dl2[c=4,1,0]=4\q
dlv2[c=4]=53\nl

T[0]=1  T[1]=1\nl  

****** TERMSCOMBINE ****\nl
th*th-term\nl

****** SORTOUTthBth ****\nl
lab50 at thBthBthth, Final result\nl 
weight= 1+i(0)\q
PlusMinus= 0\q
Sign=1\q
Nthth=1\q
NthBthB=0\q
Nhalf=0\nl

******  SigmaContraction  ****\nl
SigGraN=0\q
FinalOutPut: MultiFac=1 + i(0)\q
type2[c=2]= D\q
D2[c=2]=1\q
type2[c=3]= D\q
D2[c=3]=1\nl

T[0]=1  T[1]=2\nl  

****** TERMSCOMBINE ****\nl
th*th-term\nl

****** SORTOUTthBth ****\nl
lab50 at thBthBthth, Final result\nl 
weight= 0+i(-1)\q
PlusMinus= 0\q
Sign=0\q
Nthth=1\q
NthBthB=0\q
Nhalf=1\nl

******  SigmaContraction  ****\nl
SigGraN=201199\q
FinalOutPut: MultiFac=2 + i(0)\q
type2[c=4]= D\q
D2[c=4]=1\q
type2[c=5]= w\q
dv2[c=5]=53\q
dvv2[c=5]=53\nl

T[0]=2  T[1]=1\nl  

****** TERMSCOMBINE ****\nl
th*th-term\nl

****** SORTOUTthBth ****\nl
lab50 at thBthBthth, Final result \nl
weight= 0+i(-1)\q
PlusMinus= 0\q
Sign=1\q
Nthth=1\q
NthBthB=0\q
Nhalf=1\nl

******  SigmaContraction  ****\nl
SigGraN=201199\q
FinalOutPut: MultiFac=-2 + i(0)\q
type2[c=4]= w\q
dv2[c=4]=51\q
dvv2[c=4]=51\q
type2[c=5]= D\q
D2[c=5]=1\nl

T[0]=2  T[1]=2\nl  

****** TERMSCOMBINE ****\nl
th*th-term\nl

****** SORTOUTthBth ****\nl
lab50 at thBthBthth, Final result\nl 
weight= -1+i(0)\q
PlusMinus= 0\q
Sign=0\q
Nthth=1\q
NthBthB=0\q
Nhalf=1\nl

******  SigmaContraction  ****\nl
SigGraN=402299\nl
Branch No 0\nl
FinalOutPut: MultiFac=-2 + i(0)\q
type2[c=6]= w\q
dv2[c=6]=51\q
dvv2[c=6]=51\q
type2[c=7]= w\q
dv2[c=7]=54\q
dvv2[c=7]=54\nl
Branch No 1\nl
FinalOutPut: MultiFac=2 + i(0)\q
type2[c=6]= w\q
dv2[c=6]=52\q
dvv2[c=6]=51\q
type2[c=7]= w\q
dv2[c=7]=51\q
dvv2[c=7]=52\nl
Branch No 2\nl
FinalOutPut: MultiFac=-2 + i(0)\q
type2[c=6]= w\q
dv2[c=6]=52\q
dvv2[c=6]=51\q
type2[c=7]= w\q
dv2[c=7]=52\q
dvv2[c=7]=51\nl
Branch No 3\nl
FinalOutPut: MultiFac=0 + i(2)\q
type2[c=5]= e\q
ep(51,52,54,53)\q
type2[c=6]= w\q
dv2[c=6]=52\q
dvv2[c=6]=51\q
type2[c=7]= w\q
dv2[c=7]=54\q
dvv2[c=7]=53\nl

T[0]=3  T[1]=0\nl  

****** TERMSCOMBINE ****\nl
th*th-term\nl

****** SORTOUTthBth ****\nl
lab50 at thBthBthth, Final result\nl 
weight= 0+i(-1)\q
PlusMinus= 0\q
Sign=0\q
Nthth=1\q
NthBthB=0\q
Nhalf=0\nl

******  SigmaContraction  ****\nl
SigGraN=1\q
FinalOutPut: MultiFac=1 + i(0)\q
type2[c=2]= s\q
si2[c=2,0,1]=1\q
si2[c=2,1,1]=2\q
siv2[c=2]=51\q
type2[c=3]= m\q
dl2[c=3,1,0]=2\q
dlv2[c=3]=51\q
type2[c=4]= l\q
la2[c=4,0,0]=1\nl

Gathering all terms, we obtain
\begin{eqnarray}
W_\al W^\al=2i\graph{OutFig1r}\ {\bf \{(0,3),(3,0)\}}\ +\ 0\ {\bf (1,2)}\ +\ 0\ {\bf (2,1)}\nn
-\half\left(
0\ {\bf (2,2)br0}-\half {v_{mn}}^2\ {\bf (2,2)br1}-\half {v_{mn}}^2\ {\bf (2,2)br2}
-2i\ep^{m,n,s,l}\half v_{mn}\cdot\half v_{sl}\ {\bf (2,2)br3}\right)\nn
-D^2\ {\bf (1,1)}.
\label{WW}
\end{eqnarray}
Adding $\Wbar\Wbar$, we obtain the Lagrangian as
\begin{eqnarray}
\Lcal=\fourth (-W_\al W^\al|_{\sh^2}+\Wbar_\aldot \Wbar^\aldot|_{\thbar^2})\nn
=-\fourth {v_{mn}}^2+\frac{i}{2}(\graph{OutFig2r}-\graph{OutFig1r})+\half D^2\ ,
\label{superEM}
\end{eqnarray}
where we do not ignore the total divergence. 
This is the kinetic term of the photon and the photino.

\section*{Acknowledgements}
This work is completed in the author's stay at
ITP, Univ. Wien. He thanks A. Bartl for reading carefully the manuscript and
all menbers of the institute for the hospitality. 


\begin{thebibliography}{bib}
\bibitem{SI03}
S. Ichinose, hep-th/0301166, DAMTP-2003-8, US-03-01,
"Graphical Representation of Supersymmetry"
\bibitem{SUSY2004}
S. Ichinose, hep-th/0410027, Proc. 12th Int.Conf. on "Supersymmetry and
Unification of Fundamental Interactions"(June 17-23,2004,Epochal Tsukuba
Congress Center,Japan), p853-856, 
"Graphical Representation of Supersymmetry and Computer Calculation"
\bibitem{SI06UW07}
S. Ichinose, Univ. Vienna preprint UWThPh-2006-7, 
"Graphical Representation of Supersymmetry"
\bibitem{WB92} 
J. Wess and J. Bagger, {\it Supersymmetry and Supergravity}. 
Princeton University Press, Princeton, 1992
\bibitem{TV72}
G. 'tHooft and M. Veltman, Nucl.Phys.B44,189(1972)
\bibitem{GS85}
M.H. Goroff and A. Sagnotti, Phys.Lett.B150(1985)81;
Nucl.Phys.B266(1986)709
\bibitem{Ven92}
A.E.M. van de Ven, Nucl.Phys.B378(1992)309
\bibitem{SI98IJMPC} 
S. Ichinose, \IJMP {\bf C9}(1998)243, hep-th/9609014
%
%
%
\end{thebibliography}
\end{document}